\documentclass[flushrt]{emulateapj}

\usepackage{subfigure} 
\usepackage{epsfig}

\newcommand{\npar}{\par \vspace{2.3ex plus 0.3ex minus 0.3ex}} 
\usepackage{lscape}
\usepackage{longtable}

\begin{document}
\title{Stellar Populations of Highly Magnified Lensed Galaxies: Young Starbursts at $z\sim2$.} 

\author{Eva Wuyts\altaffilmark{1,2}, Jane R. Rigby\altaffilmark{3}, Michael D. Gladders\altaffilmark{1,2}, David G. Gilbank\altaffilmark{4}, Keren Sharon\altaffilmark{2}, Megan B. Gralla\altaffilmark{1,2}, Matthew B. Bayliss\altaffilmark{1,2}}

\footnotetext[*]{Based in part on observations collected at the 3.5\,m Apache Point Observatory telescope in New Mexico, which is owned and operated by the Astrophysical Research Consortium}
\altaffiltext{1}{Department of Astronomy and Astrophysics, University of Chicago, 5640 S. Ellis Av., Chicago, IL 60637}
\altaffiltext{2}{Kavli Institute for Cosmological Physics, University of Chicago, 5640 South Ellis Avenue, Chicago, IL 60637}
\altaffiltext{3}{Observational Cosmology Lab, NASA Goddard Space Flight Center, Greenbelt, MD 20771}  
\altaffiltext{4}{Department of Physics and Astronomy, University of Waterloo, Waterloo, Ontario, N2L 3G1, Canada}

\begin{abstract}
We present a comprehensive analysis of the rest-frame UV to near-IR spectral energy distributions and rest-frame optical spectra of four of the brightest gravitationally lensed galaxies in the literature: RCSGA 032727-132609 at $z=1.70$, MS1512-cB58 at $z=2.73$, SGAS J152745.1+065219 at $z=2.76$ and SGAS J122651.3+215220 at $z=2.92$. This includes new Spitzer imaging for RCSGA0327 as well as new spectra, near-IR imaging and Spitzer imaging for SGAS1527 and SGAS1226. Lensing magnifications of 3-4 magnitudes allow a detailed study of the stellar populations and physical conditions. We compare star formation rates as measured from the SED fit, the H$\alpha$ and [O~II]~$\lambda$3727 emission lines, and the UV+IR bolometric luminosity where 24\,$\mu$m photometry is available. The SFR estimate from the SED fit is consistently higher than the other indicators, which suggests that the Calzetti dust extinction law used in the SED fitting is too flat for young star-forming galaxies at $z\sim2$. Our analysis finds similar stellar population parameters for all four lensed galaxies: stellar masses $3-7 \times 10^9$\,M$_\odot$, young ages $\sim$ 100\,Myr, little dust content $E(B-V)$=0.10-0.25, and star formation rates around 20-100\,M$_\odot$~yr$^{-1}$. Compared to typical values for the galaxy population at $z\sim2$, this suggests we are looking at newly formed, starbursting systems that have only recently started the build-up of stellar mass. These results constitute the first detailed, uniform analysis of a sample of the growing number of strongly lensed galaxies known at $z\sim2$. 

\subjectheadings{galaxies: high-redshift, strong gravitational lensing, infrared: galaxies}                                    
\end{abstract}

\section{Introduction}
Recent years have seen a significant rise in the number of known gravitationally lensed galaxies at $z=1-3$, both reported from well-defined survey searches (Bolton et al. 2006, Cabanac et al. 2007, Hennawi et al. 2008, Lin et al. 2009, Gladders et al. 2011 in preparation) and from serendipitous discoveries \citep{Allam:07,Belokurov:07,Smail:07}. The lensing magnification grants access to high signal-to-noise photometric and spectroscopic observations over a wide range of wavelengths and creates unique opportunities for very detailed study of the stellar populations, dynamics and physical conditions of star-forming galaxies at $z\sim2$, complementing statistical studies of large samples of unlensed star-forming galaxies at these redshifts (e.g. Shapley et al. 2005, Erb et al. 2006a, Daddi et al. 2007, Reddy et al. 2010). Today, these studies are limited by systematic uncertainties related to the recipes used to estimate physical parameters such as metallicity, dust extinction, star formation rate and star formation history. Until the era of extremely large telescopes, lensed galaxies provide unique opportunities to increase our physical understanding of the composition and assembly history of stellar populations during this period of peak star formation in the Universe. 

So far most lensing systems have been studied individually, based on different follow-up observations and analysis techniques. To take advantage of the growing number of known strongly lensed sources, it is crucial to invest in a uniform study of their properties.
In this paper we present a full analysis of multi-wavelength photometry and rest-frame optical spectroscopy of four of the brightest distant lensed galaxies known to date. RCSGA 032727-132609 was recently discovered in the Second Red-Sequence Cluster Survey \citep{Gilbank:11}; initial analysis of optical/near-IR photometry and construction of a lens model for the foreground galaxy cluster are described in \cite{Wuyts:10}. Rest-frame optical spectroscopy from Keck/NIRSPEC is presented in \cite{Rigby:11}. This paper adds Spitzer imaging to the SED to probe the rest-frame near-IR. MS1512-cB58 has been studied extensively since its discovery \citep{Yee:96} and a multitude of photometric and spectroscopic data is available for this system \citep{Ellingson:96, Teplitz:00, Siana:08}. SGAS J152745.1+065219 and SGAS J122651.3+215220 are part of the SDSS Giant Arcs Survey \citep{Bayliss:11}; optical photometry and lens models are reported in \cite{Koester:10}. Here we present near-IR and Spitzer imaging as well as rest-frame optical spectra for both sources. We will refer to these galaxies as RCSGA0327, cB58, SGAS1527 and SGAS1226.

The paper is organized as follows. \S\ref{sec:data} describes the near-IR and Spitzer imaging and the rest-frame optical spectroscopy. The methods used for the photometric analysis and spectral energy distribution modeling, as well as the various reddening and star formation rate diagnostics are discussed in \S\ref{sec:meth}. \S\ref{sec:results} presents the physical conditions and stellar populations of the lensed galaxies as derived from these methods. We adopt a flat cosmology with $\Omega_M = 0.3$ and H$_0 = 70$\,km\,s$^{-1}$\,Mpc$^{-1}$. All magnitudes are quoted in the AB system.

\section{Observations and Data Reduction}
\label{sec:data}
\subsection{Imaging}
\label{sec:im}
Optical and near-IR photometry of RCSGA0327 and optical photometry of SGAS1527 and SGAS1226 are taken from their respective discovery papers \citep{Wuyts:10, Koester:10}. For cB58, we take optical and near-IR photometry from \cite{Ellingson:96} and Spitzer photometry from \cite{Siana:08}. Here we present new Spitzer imaging for RCSGA0327 as well as new near-IR and Spitzer imaging for SGAS1527 and SGAS1226.
\npar
We observed SGAS1527  and SGAS1226 with the 3.5\,m telescope at the Apache Point Observatory (APO) in New Mexico in the $J$-, $H$- and $K_s$-bands during four half nights on 2010 January 30, February 24, March 4 and March 29 with seeing of $0\farcs7$-$0\farcs9$. A pipeline of standard IRAF\footnotemark[1] tasks is used to sky-subtract and stack the dithered images, giving total integration times of 3600\,s - 4500\,s - 3600\,s in $J$ - $H$ - $K_s$ for SGAS1527 and 3600\,s - 3150\,s - 2700\,s in $J$ - $H$ - $K_s$ for SGAS1226. 

IRAC \citep{Fazio:04} and MIPS \citep{Rieke:04} observations of RCSGA0327 and SGAS1527 were obtained through Spitzer program 50823 (PI: M.~D. Gladders). RCSGA0327 was observed for 600\,s in all four IRAC bands (3.6/4.5/5.8/8.0\,$\mu$m) and 700\,s in MIPS 24\,$\mu$m; SGAS1527 was observed for 3000\,s in all IRAC bands and 1300\,s at 24\,$\mu$m. 600\,s warm IRAC observations of SGAS1226 at 3.6 and 4.5\,$\mu$m were obtained through Spitzer program 70154 (PI: M.~D. Gladders). The data is reduced with the MOPEX software distributed by the Spitzer Science Center (SCC) and drizzled to a finer pixel scale of $0\farcs5$\,pix$^{-1}$ for IRAC and $1\farcs5$\,pix$^{-1}$ for MIPS. 

\footnotetext[1]{IRAF (Image Reduction and Analysis Facility) is distributed by the National Optical Astronomy Observatories, which are operated by AURA, Inc., under cooperative agreement with the National Science Foundation.}

\subsection{Rest-frame optical spectroscopy}
\label{sec:spec}
\begin{figure*}
\centering
\includegraphics[width=8cm]{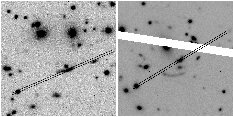}
\vspace{0.3in}
\caption{Finder charts showing NIRSPEC slit positions for SGAS1527 \textit{(left)} and SGAS1226 \textit{(right)}, both $45\times45$\arcsec. The CFHT Megacam $r$-band image of SGAS1527 and the Gemini GMOS $i$-band of SGAS1226 \citep{Koester:10} are overplotted with the NIRSPEC 0.76\arcsec\ $\times$ 42\arcsec\ longslit at position angles of 112\arcdeg\ and 121\arcdeg East of North respectively. The missing region in the $i$-band image of SGAS1226 is due to the GMOS chip gap. \label{fig:findercharts}}
\end{figure*}

Keck/NIRSPEC rest-frame optical spectra were published for cB58 by \cite{Teplitz:00} and for RCSGA0327 by \cite{Rigby:11}. Here we present new spectra for SGAS1527 and SGAS1226.
\npar
We observed SGAS1527 and SGAS1226 with the NIRSPEC spectrograph \citep{McLean:98} on the Keck II telescope on the night of 2010, February 4. The weather was clear, and the seeing was measured as 0.85\arcsec\ and 0.45\arcsec\ when the telescope was focused during the night. We used the low-resolution mode and the 0.76\arcsec\ $\times$ 42\arcsec\ slit which was rotated to position angles of 112\arcdeg\ and 121\arcdeg\ East of North, respectively, to maximize coverage of each arc. Finder charts showing the location of the slit on each target are given in Figure~\ref{fig:findercharts}. The targets were acquired by offsetting from the brightest cluster galaxy on the near-IR slit-viewing camera; target acquisition was verified by direct imaging with this camera. The targets were nodded along the slit in an AB pattern, with exposures of 600\,s. Table~\ref{tab:obslog} summarizes the filters and total integration times. For SGAS1527, the default grating angles were used, but for SGAS1226 the grating angles were adjusted to give the wavelength ranges reported. The A0V stars 7 Ser and HD 109055 were observed every hour as telluric standards for SGAS1527 and SGAS1226 respectively. In addition, V.~Tilvi and J.~Rhoads of the Arizona State University have kindly obtained additional spectra of SGAS1527 on 2010, September 17, with the same setup. The AOV star HD 12021, observed 10~hr later, is used as a telluric standard. Due to the time delay, the absolute fluxing in this bandpass should not be trusted.

\begin{deluxetable}{cccc}
\tablecolumns{4} 
\tablecaption{Spectroscopic Observation Log.}
\tablehead{ \multicolumn{1}{c}{source} &
	    \multicolumn{1}{c}{filter}   & 
            \multicolumn{1}{c}{time}  & 
	    \multicolumn{1}{c}{wavelength range} \\
	    \multicolumn{1}{c}{} &
	    \multicolumn{1}{c}{}   & 
            \multicolumn{1}{c}{s}  & 
	    \multicolumn{1}{c}{$\mu$m}}
\startdata

SGAS1527   & NIRSPEC-4     & 1800  &  1.302--1.585 \\
     	   & NIRSPEC-6     & 3000  &  1.770--2.190 \\
           & NIRSPEC-7     & 1620  &  2.242--2.521 \\
\enddata

\startdata
SGAS1226   & NIRSPEC-5b    & 1800   &  1.447--1.730 \\
           & NIRSPEC-7b    & 3000   &  1.864--2.285 \\
\enddata
\label{tab:obslog}
\end{deluxetable}

\begin{figure*}
\centering
\includegraphics[width=2.2in]{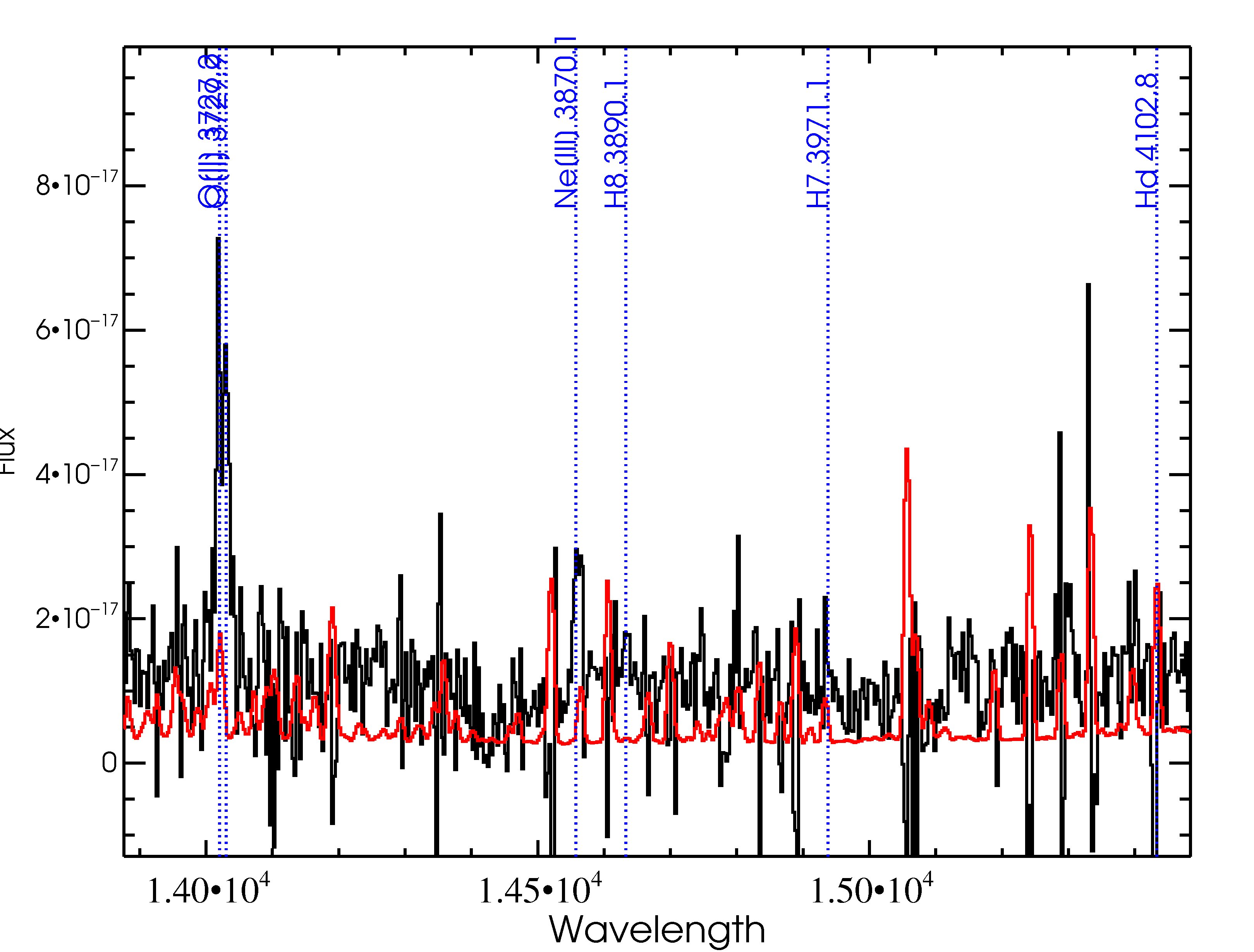}
\includegraphics[width=2.2in]{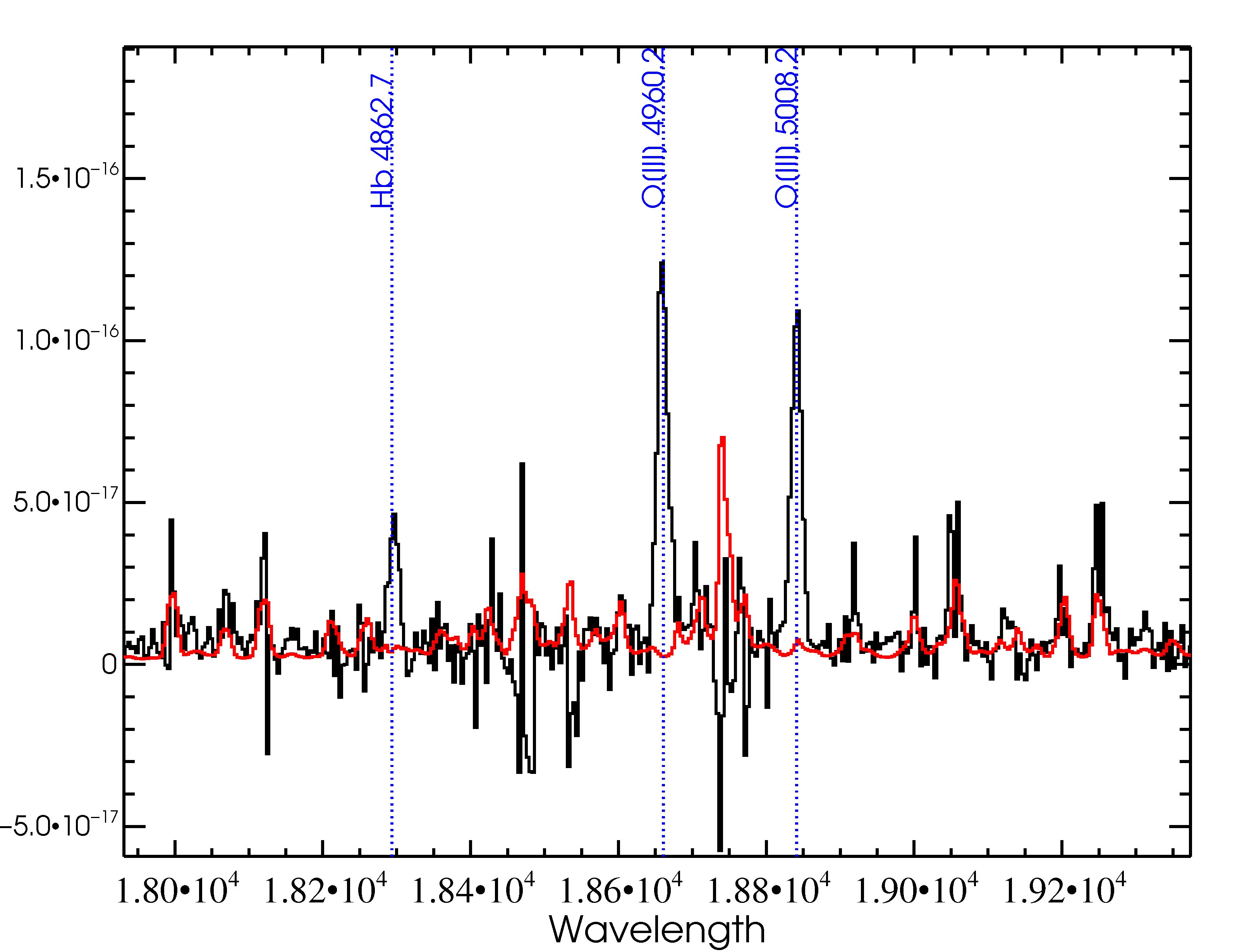} 
\includegraphics[width=2.2in]{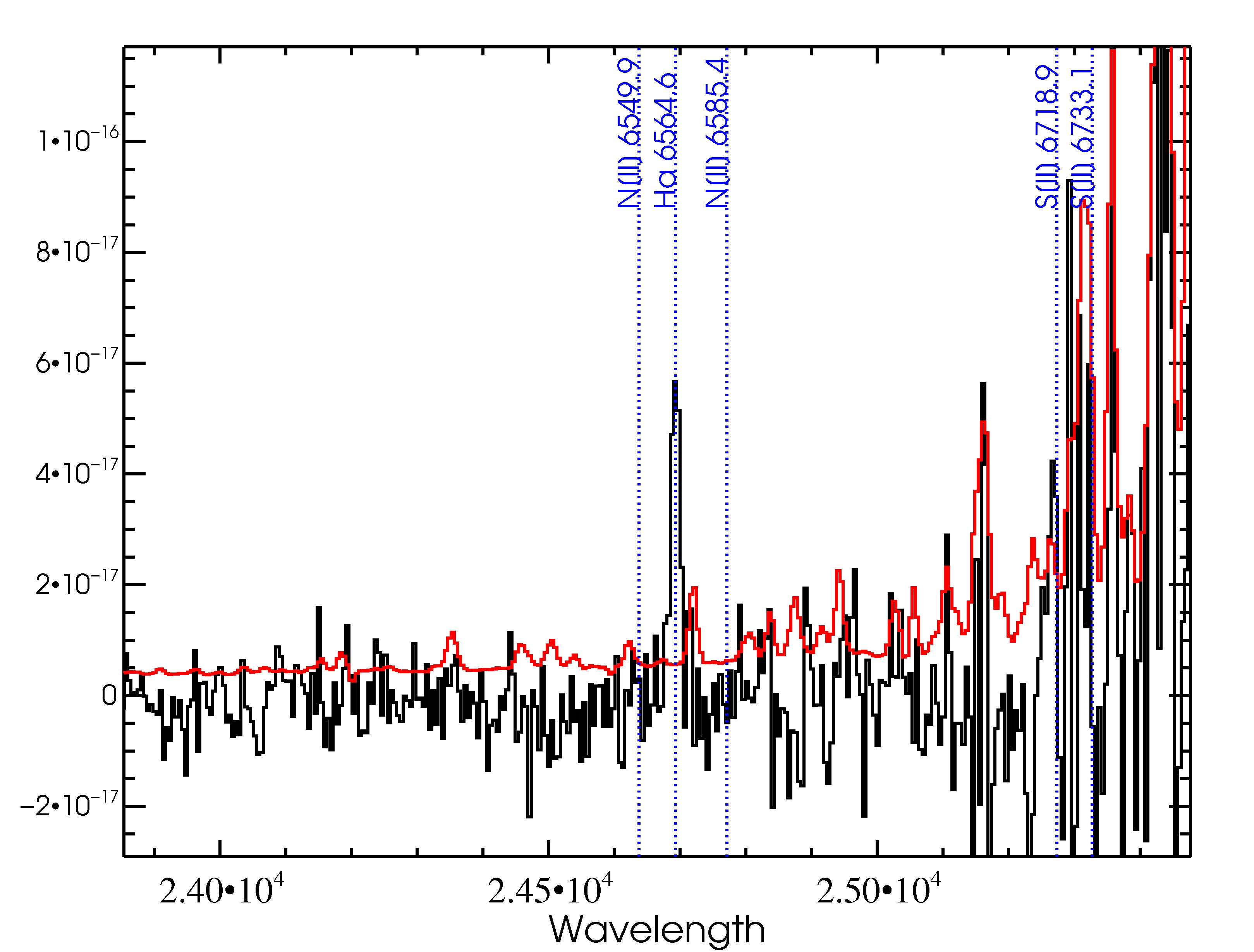} 
\figcaption{NIRSPEC spectra for SGAS1527. The fluxed spectra are plotted in black, with the $1\sigma$ error spectrum in red. The X-axis shows observed wavelength in Angstroms; the Y-axis shows observed flux in units of erg~s$^{-1}$~cm$^{-2}$. Blue labels mark detected emission lines as well as lines for which we measure upper limits. \label{fig:spec_1527}}
\end{figure*}

\begin{figure*}
\begin{center}
\includegraphics[width=2.2in]{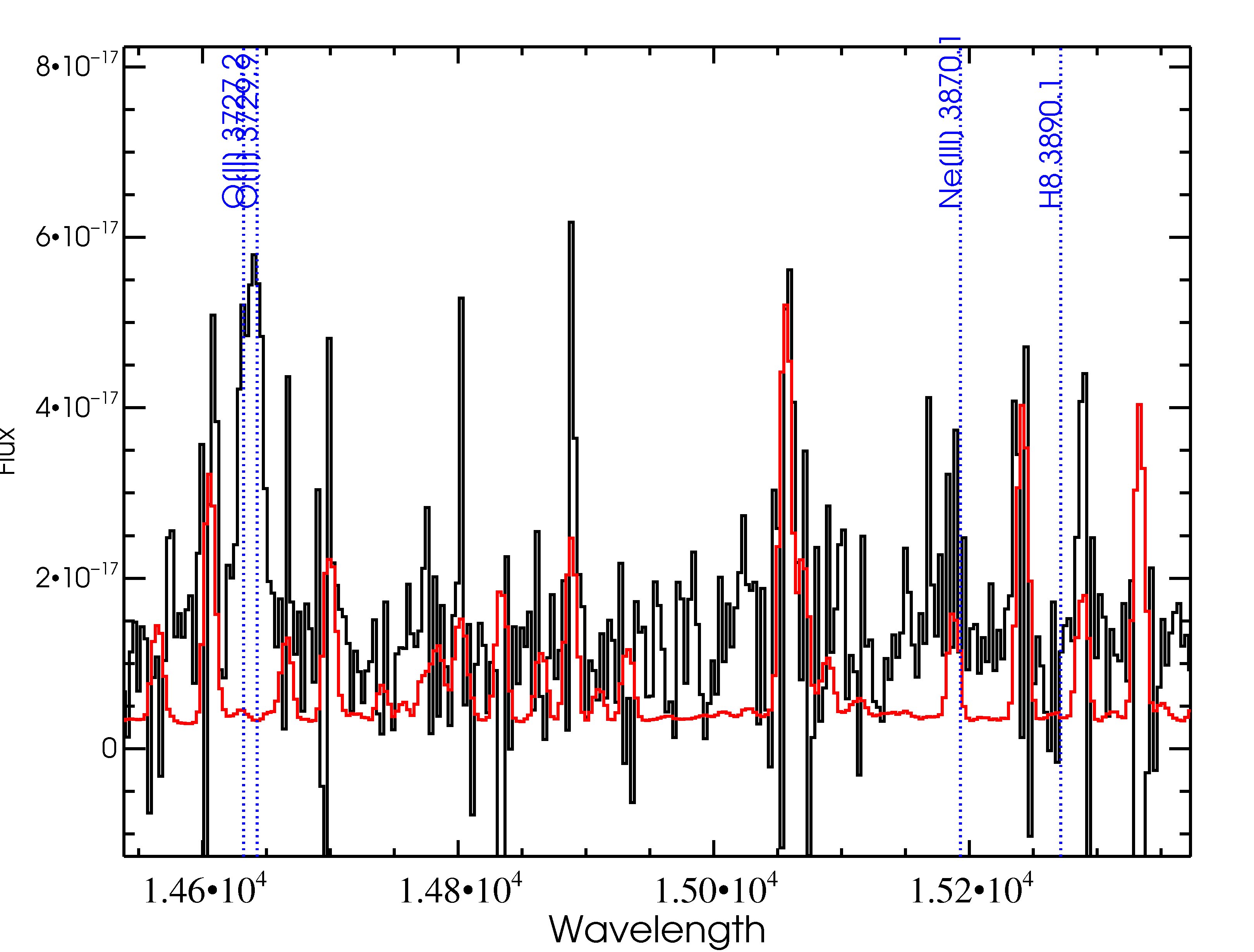} 
\includegraphics[width=2.2in]{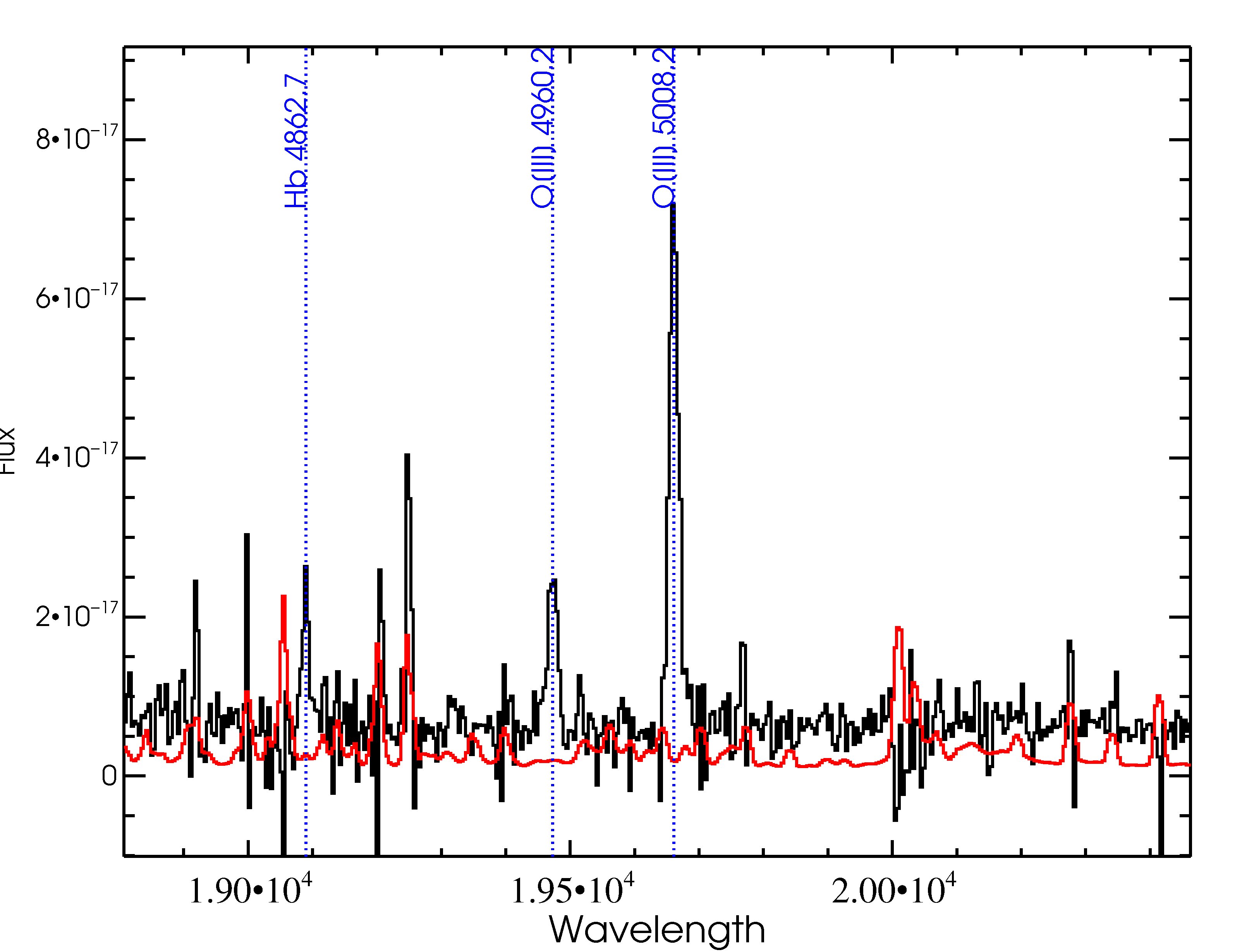} 
\figcaption{NIRSPEC spectra for SGAS1226. The fluxed spectra are plotted in black, with the $1\sigma$ error spectrum in red. The X-axis shows observed wavelength in Angstroms; the Y-axis shows observed flux in units of erg~s$^{-1}$~cm$^{-2}$. Blue labels mark detected emission lines as well as lines for which we measure upper limits.\label{fig:spec_1226}}
\end{center}
\end{figure*}

The spectra are reduced with the \textit{nirspec$\_$reduce} package written by G.~D.~Becker. The data reduction, fluxing, and line-fitting procedures are described in \citet{Rigby:11}. Solar abundances are taken from Table~1 of \citet{Asplund:09}. The reduced spectra are plotted in Figures~\ref{fig:spec_1527} and \ref{fig:spec_1226}; line fluxes are reported in Table~\ref{tab:fluxes}. At observed wavelengths where the Earth's atmosphere has very little transmission, small mismatches in airmass and precipitable water vapor (PWV) between the telluric and the science target can have a large effect on the flux calibration. To quantify this, Table~\ref{tab:fluxes} reports the expected atmospheric transmission\footnotemark[2] for each detected emission line integrated over the gaussian line width as given in \S\ref{sec:resultsspec}. The quoted uncertainty combines the measurement uncertainty in the line widths, the systematic uncertainty in PWV and the systematic uncertainty from a 0.1 airmass mismatch between the telluric and science target. Subsequent calculations involving line fluxes will take into account both the 1$\sigma$ flux uncertainties and the fractional uncertainty on the atmospheric transmission.
\footnotetext[2]{The atmospheric transmission is based on the model absorption spectra for Mauna Kea published by the Gemini Observatory: http://www.gemini.edu/sciops/telescopes-and-sites/observing-condition-constraints/ir-transmission-spectra. The PWV at the time of observations was derived from historical atmospheric opacity data at http://puuoo.caltech.edu/.}

\begin{deluxetable}{cccc}
\tablecolumns{4} 
\tablecaption{Measured line fluxes.} 
\tablehead{ \multicolumn{1}{c}{line ID} &
	    \multicolumn{1}{c}{$\lambda_{obs}$}   & 
            \multicolumn{1}{c}{flux}  & 
	    \multicolumn{1}{c}{atm trans} \\
            \multicolumn{1}{c}{} &
	    \multicolumn{1}{c}{$\mu$m}   & 
            \multicolumn{1}{c}{$10^{-16}$~erg~s$^{-1}$~cm$^{-2}$}  & 
	    \multicolumn{1}{c}{}}

\startdata 
\multicolumn{4}{c}{\textbf{SGAS1527 - Filter N4}} \\
$\mathrm{[OII]~3727}$   & 1.40208 & 11$\pm$1     & 0.54$\pm$0.08 \\
$\mathrm{[NeIII]~3869}$ & 1.45578 & 4$\pm$1      & 0.62$\pm$0.07 \\
                        &         &              &               \\
\multicolumn{4}{c}{\textbf{SGAS1527 - Filter N6}} \\
H$\beta$                & 1.82932 & 9$\pm$2      & 0.17$\pm$0.08 \\
$\mathrm{[OIII]~4959}$  & 1.86616 & 20.0$\pm$0.2 & 0.23$\pm$0.08 \\
$\mathrm{[OIII]~5007}$  & 1.88397 & 18.0$\pm$0.3 & 0.64$\pm$0.08 \\
                        &         &              &               \\
\multicolumn{4}{c}{\textbf{SGAS1527 - Filter N7}} \\
\vspace{0.1in}
H$\alpha$               & 2.46925 & 10$\pm$3     & 0.82$\pm$0.03 \\
                        &         &              &               \\
\multicolumn{4}{c}{\textbf{SGAS1226 - Filter N5b}} \\
$\mathrm{[OII]~3727}$   & -----   & 9.6$\pm$0.7  & 0.76$\pm$0.06 \\
                        &         &              &               \\
\multicolumn{4}{c}{\textbf{SGAS1226 - Filter N7b}} \\
H$\beta$                & 1.90895 & 1.8$\pm$0.4  & 0.55$\pm$0.09 \\
$\mathrm{[OIII]~4959}$  & 1.94697 & 4.5$\pm$0.7  & 0.74$\pm$0.04 \\
$\mathrm{[OIII]~5007}$  & 1.96641 & 13$\pm$4     & 0.67$\pm$0.03 \\
\enddata

\label{tab:fluxes}
\end{deluxetable}


\section{Methods}
\label{sec:meth}
\subsection{Photometry}
\label{sec:phot}
The photometric analysis of the imaging data follows the method detailed in \cite{Wuyts:10}. In short, the data are transformed to a common reference and pixel scale and empirical, normalized point spread functions (PSF) are created for each image. We define object apertures by tracing a curve along the extended source and convolving it with the appropriate PSF. A series of apertures of increasing radial extent are defined as isophotes of this convolution. The apertures can be very much non-circular and are described by equivalent radii based on circular apertures that extend to the same isophotes. We use the GALFIT package (version 3.0, Peng et al. 2010) to fit Sersic profiles to neighboring galaxies that fall within the object apertures and subtract these from the images. Accurate masking of neighboring galaxies becomes very important for the Spitzer data due to the extended, non-circular PSFs of the IRAC and MIPS instruments. We have found the most robust masking method to be based on cross-convolution. This consists of convolving the Spitzer data with the PSF of a chosen optical band and similarly convolving the optical GALFIT models of neighboring sources with the relevant Spitzer PSF. These convolved models are scaled as needed and subtracted from the convolved Spitzer image. 

We measure magnitudes at an equivalent radius of twice the FWHM of the image. The optical and near-IR data are aperture-corrected to an equivalent radius of $6''$ based on the curve of growth of the PSF reference star. In \cite{Wuyts:10}, we convolved all optical and near-IR images of RCSGA0327 to the spatial resolution of the $H$-band, which presents the worst seeing over the 9 optical/near-IR bands, in order to measure flux from the same physical region of the source in all bands. This is not repeated here; by aperture correcting final magnitudes to an equivalent radius of $6''$, the total source flux in each band is measured, regardless of prior PSF-matching. We have reanalyzed the data for RCSGA0327 without matching the PSFs and find final magnitudes consistent with the photometry reported in \cite{Wuyts:10}, within the $1\sigma$ uncertainties. The aperture corrections for the IRAC data are applied as detailed in the IRAC handbook\footnotemark[3], the correction for the MIPS data is calculated from a Tiny Tim model of the 24\,$\mu$m PSF. The MIPS photometry also requires a color correction of -0.04\,mag, as stated in the MIPS handbook\footnotemark[4] (assuming $f_\nu \sim \nu^{-2}$). We derive limiting magnitudes for a few non-detections in the Spitzer data: the counter-image of RCSGA0327 is not detected at 5.8 and 8.0\,$\mu$m, and SGAS1527 is not detected at 24\,$\mu$m. We convolve the image of the source in a chosen optical band with the relevant Spitzer PSF, scale it and manually insert it into the convolved Spitzer image at 100 random positions away from real sources. Limiting magnitudes are determined from the scaling where SExtractor \citep{sex} detects 90\% of these mock objects at $3\sigma$.
\footnotetext[3]{http://ssc.spitzer.caltech.edu/irac/iracinstrumenthandbook/}
\footnotetext[4]{http://ssc.spitzer.caltech.edu/mips/mipsinstrumenthandbook/}
Final magnitudes are corrected for galactic extinction \citep{Schlegel:98}. Photometric uncertainties include Poisson noise, absolute zeropoint uncertainties, uncertainties from the Spitzer aperture corrections as detailed in the instrument handbooks, and uncertainties from the masking of neighboring galaxies; the latter two dominate the total uncertainty.
We independently analyze the IRAC and MIPS data for cB58 and find final magnitudes consistent with the photometry in \cite{Siana:08}. We choose to use uncertainties twice as large as reported there, to take into account the influence of the halo of the neighboring cD galaxy.

\subsection{Spectral energy distribution modeling}
\label{sec:methsed}
The available photometry fully characterizes the rest-frame UV to near-IR spectral energy distributions of these galaxies, allowing us to constrain the stellar populations (stellar mass, age, extinction and star formation rate) through a comparison to stellar population synthesis models. The presence of strong emission lines at rest-frame optical wavelengths could affect the broad-band photometry and therefore the SED fit. From the emission line fluxes derived from the rest-frame optical spectroscopy, we estimate the contribution of the strongest emission lines to the $H$- and $K_s$-band magnitudes to be at most 5-10\%, which does not have a significant effect compared to the estimated photometric errors. \cite{Reddy:10} have also shown that the effect of emission lines on the best-fit age and stellar mass is further reduced when IRAC data (which is unaffected by line contamination at these redshifts) is incorporated. 

The fitting procedure, as well as its caveats and degeneracies, is explained in more detail in \cite{Wuyts:10}. We use \textit{Hyperz} \citep{Bolzonella:00} to perform SED fitting at a fixed spectroscopic redshift. The newest Bruzual \& Charlot population synthesis models (CB07, kindly made available by the authors - see \cite{Bruzual:03}) are used, with a Chabrier initial mass function \citep{Chabrier:03} and a Calzetti dust extinction law \citep{Calzetti:00}. To reduce the number of degrees of freedom and accompanying degeneracies, we fix the metallicity at 0.4\,Z$_\odot$ for all 4 sources, consistent with the abundances indicated by rest-frame optical spectroscopy (see \S\ref{sec:resultsspec}). 
SED templates in the literature most often use constant star formation models or exponentially declining models - SFR $\sim$ e$^{-t/\tau}$ - with a wide range of values for the \textit{e}-folding time $\tau$. The current SFR reported by the SED fitting procedure depends strongly on the assumed star formation history. Other stellar population parameters like dust extinction, age and even stellar mass can be influenced as well; they all work together to produce the observed SED and numerous degeneracies and trade-offs exist. The range of SFHs allowed in the SED fit is therefore not an arbritrary choice. Exponentially declining models assume that we observe the galaxy at its minimum SFR, which is not necessarily justified for star-forming galaxies at z$\sim$2. \cite{Maraston:10} explore inverted $\tau$-models where the star formation rate rises exponentially with time - SFR $\sim$ e$^{+t/\tau}$ - and find a significant improvement in the reproduction of stellar population parameters for a sample of mock star-forming galaxies at $z \sim 2$. For this sample of lensed galaxies, we choose to limit ourselves to a constant star formation history (CSF) as a reasonable average over the galaxy's lifetime and a compromise between exponentially declining and rising SFHs. \cite{Erb:06c} also find the current SFR to be an adequate representation of the past average SFR for their sample of UV-selected galaxies at $z\sim2$. Additionally, CSF models allow a more robust comparison to other SFR indicators discussed in \S\ref{sec:methsfr}, for which the conversions from luminosity are generally based on constant star formation stellar population models. 

\begin{figure}[h]
\centering
\includegraphics{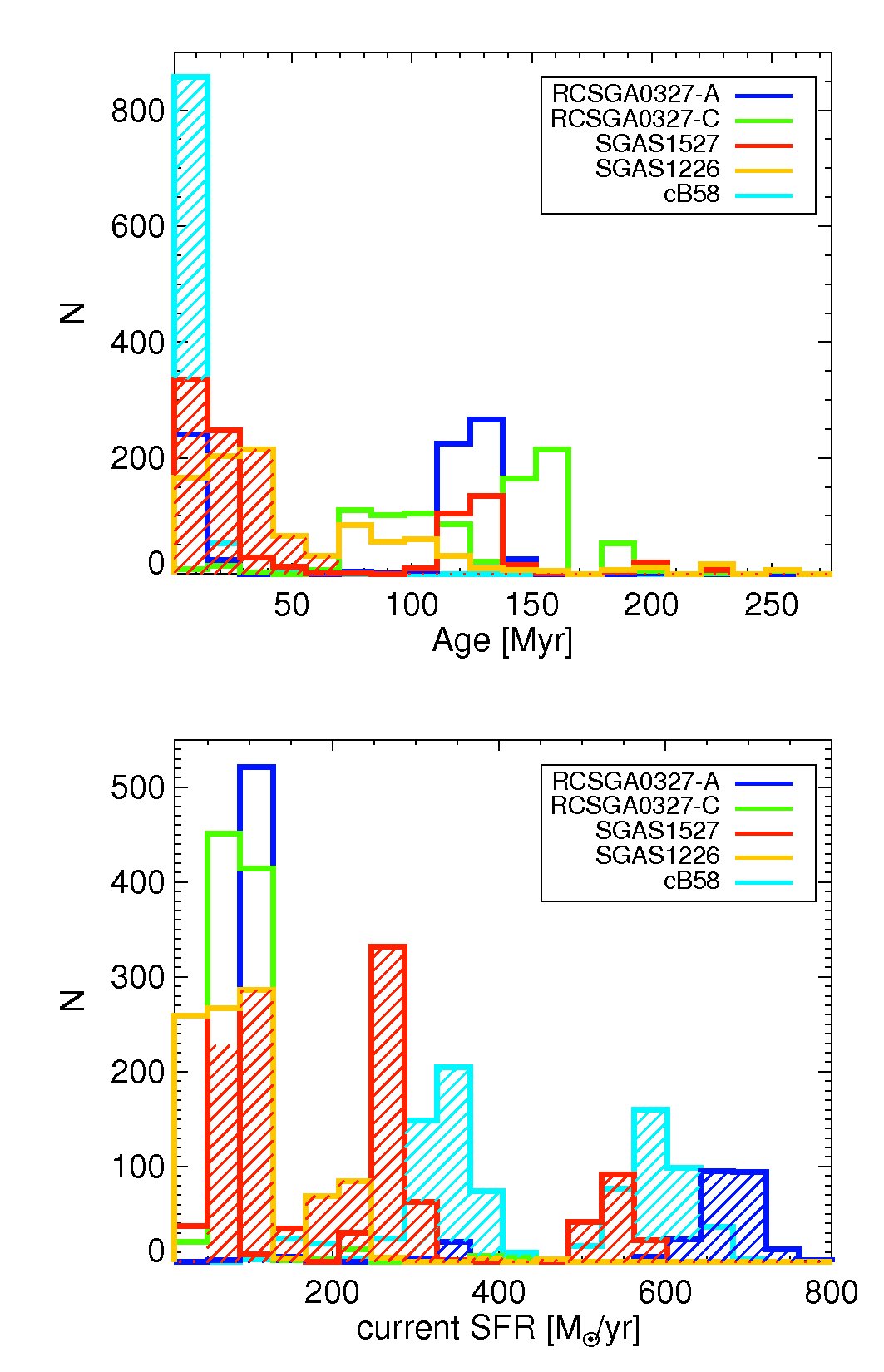}
\caption{Histograms of the best-fit age \textit{(top)} and SFR \textit{(bottom)} for 1000 mock SEDs consistent with the photometric uncertainties. A clear bimodality can be seen. The shaded regions show the best-fit SED models with ages $<70$\,Myr, which correspond exclusively to the high SFR solutions. This part of parameter space is eliminated when the age is restricted to $\ge$ 70\,Myr. \label{fig:agesfrhist}}
\end{figure}

\npar
Now that we have fully defined the input SED models, we have to consider the allowed range of stellar population parameters, specifically the stellar age. The dominant emission of O and B stars in the observed SED of a star-forming galaxy can cause the inferred stellar population parameters to be luminosity-weighted towards very young models, with unrealistically high SFRs and low stellar masses. An often used measure to avoid this luminosity-weighted bias restricts the age of the stellar population to be larger than the dynamical timescale of the galaxy \citep{Wuyts:07, Maraston:10, Reddy:10}. From velocity dispersion and size measurements of $z\sim2$ LBGs, this is inferred to be $\sim 70$\,Myr \citep{Erb:06c}.   
We illustrate this bias with histograms of the best-fit ages and star formation rates reported for 1000 mock realizations of the observed SEDs consistent with the photometric uncertainties (Figure~\ref{fig:agesfrhist}). There is a clear bimodality of very young models ($\le 70$\,Myr) and moderately older models (70-200\,Myr), which is correlated with a bimodality in the SFR, where the young models require a much higher current SFR to build up a similar stellar mass. Restricting the age to be larger than the dynamical timescale avoids this class of unphysically young stellar populations with extreme SFRs. For cB58, only very young models (ages $<$ 20~Myr) are found for the mock SEDs, due to the lack of a significant Balmer break between the $J$ and $H$-band photometry from \cite{Ellingson:96}. This was also found by \cite{Siana:08}, who report a best fit age of $9.3^{+4.7}_{-3.1}$~Myr. However, when the age is restricted to $\ge 70$~Myr, more than 90~\% of the mock SEDs return a best-fit model with $\chi^2 < 5.0$, which is not unreasonable. We will also show in \S\ref{sec:resultssfr} that the extreme SFRs which accompany very young best-fit models for cB58 (300-600~M$_\odot$~yr$^{-1}$ as can be seen from Figure~\ref{fig:agesfrhist}) completely disagree with the other SFR indicators.

A further argument in favor of the age restriction comes from the metallicity. When the metallicity is allowed to vary between 0.2\,Z$_\odot$, 0.4\,Z$_\odot$ and Z$_\odot$, the SED fitting procedure favors a metallicity of 0.2\,Z$_\odot$ for all 4 galaxies, lower than indicated by observations of the rest-frame optical spectra. When the age is forced to be $\ge 70$\,Myr, a metallicity of 0.4\,Z$_\odot$ is favored, consistent with the results from rest-frame optical spectroscopy.

\npar
Restricting the age of the stellar population limits the sensitivity of the SED fit to the most recently formed stars (on $<70$\,Myr timescales) and returns more physically meaningful stellar population parameters. A related issue is the presence of an old underlying stellar population (1-2\,Gyr timescales), which can contain the majority of a galaxy's stellar mass without emitting enough light to be detected in the observed SED at rest-frame UV and optical wavelengths. The availability of IRAC data at rest-frame near-IR wavelengths greatly increases the leverage, but some contribution of old stars cannot be ruled out from single component SED fitting. Full-scale multi-component SED fitting can address the presence of older stellar populations, but it introduces additional degrees of freedom (the number, relative timing and relative strength of the different star formation episodes), which the data fail to constrain uniquely. We can set an upper limit on the contribution of an old stellar population to the total stellar mass of the galaxies with the extreme case of a very young model, minimally reddened to match the UV spectral slope, combined with a maximally old underlying model (Daddi et al. 2004, Shapley et al. 2005, Wuyts et al. 2010). Specifically, we scale a 10\,Myr CSF model to the observed $r$-band magnitude, subtract this model from the observed SED, and scale the maximally old model ($t=3$\,Gyr for RCSGA0327 and $t=2$\,Gyr for SGAS1527, SGAS1226 and cB58) to match the residual magnitude at 3.6\,$\mu$m. In this scenario, the stellar mass is largely dominated by the older stellar population, but we find it to be at maximum a factor of 1.7 higher than the single-component mass for all sources (we have corrected the mass of the older stellar population for the mass returned to the ISM by supernovae, this is negligible for ages $\sim$10-100\,Myr \citep{Renziniciotti:93}). This places a conservative constraint on the maximum contribution of old stellar populations to the total stellar mass in these galaxies. 

\subsection{Reddening diagnostics}
\label{sec:methav}
The presence of dust in galaxies causes reddening. Since the stars and the ionized gas are not necessarily found in the same locations within a galaxy, the stellar light and the nebular emission lines may not experience the same reddening. In local star-forming galaxies, \cite{Calzetti:94} found the reddening of the nebular lines to be a factor of $\sim2$ higher than the reddening of the stellar continuum. At $z\sim2$, the same effect has been seen for some star-forming galaxies, both lensed and non-lensed \citep{Forsterschreiber:09, Finkelstein:09, Bian:10}, but other galaxies show no reddening difference between the stars and ionized gas \citep{Erb:06c,Hainline:09}.
\npar
The SED fitting procedure reports $E(B-V)_s$, the reddening of the stellar light. We add the subscript to differentiate with the reddening experienced by the galaxy's ionized gas, which we will denote as $E(B-V)_g$. 
This reddening can be derived from the Balmer decrement, the ratio of the H$\alpha$ to H$\beta$ flux. We assume an intrinsic Balmer decrement of 2.85, obtained for recombination case B, electron density $n_e \le 10^4$\,cm$^3$ and temperature $T \sim 10^4$\,K \citep{Osterbrock:89}. 
When no measure of the Balmer decrement is available, we use the empirical relation between a galaxy's stellar mass and its gas reddening derived at $z\sim0.1$ from the Stripe 82 subsample of the Sloan Digital Sky Survey (Eq. 9, Gilbank et. al. 2010). This G10 relation has an uncertainty of $\sim 0.2$~dex on $E(B-V)_g$ at the relevant stellar masses \footnotemark[5]. This empirical relation has not been verified directly at higher redshift, although agreements with 24~$\mu$m emission have been found at $z\sim1$ \citep{Gilbank:10}, but it is clearly preferable to assuming an intrinsic Balmer decrement.
\footnotetext[5]{\cite{Gilbank:10} use the Kroupa IMF, we convert to the Chabrier IMF with $\log(\mathrm{M}_*)_{\mathrm{Chab}}=\log(\mathrm{M}_*)_{\mathrm{Kroupa}}-0.04$}

\subsection{Star formation rate diagnostics}
\label{sec:methsfr}
A galaxy's star formation rate can be estimated from several different indicators, the most common of which are the UV continuum emission, sometimes combined with the infrared emission from dust grains heated by absorbed UV photons, and the nebular emission lines from the ionized gas. Rest-frame UV luminosity is straightforward to measure at $z\sim2$, but the conversion to a SFR depends on the galaxy's stellar population and the luminosity requires a large correction for dust obscuration. 
Dust corrections can be avoided by including the galaxy's infrared emission, which traces the stellar light absorbed by dust grains, such that the combined UV and IR luminosity captures the total bolometric output from young stars.

Rest-frame optical emission line indicators require a smaller dust correction than the UV continuum, but need to be extrapolated from the regions of the galaxy sampled by the spectroscopic aperture. These aperture corrections are especially important for lensed galaxies, where the arcs are often too extended to be covered by the slit and a precise magnification model is required to extrapolate the observed line emission to the total intrinsic emission of the galaxy. Additionally, in the presence of strong stellar population gradients, the observed region might not be representative of the galaxy as a whole.

\subsubsection{Star formation from UV luminosity}
\label{sec:sfruv}
Star formation rates based on the UV continuum are easy to obtain at $z\sim2$. Two or more broadband magnitudes at rest-frame UV wavelengths suffice to measure the UV luminosity and to estimate the dust correction from the UV spectral slope, as established by \cite{Meurer:99} for local starburst galaxies (based on the Calzetti extinction law). However, this method involves considerable uncertainties. The conversion between UV luminosity and SFR depends strongly on the assumed stellar population model, mostly its age and star formation history, causing an uncertainty of at least 0.3~dex in the calibration \citep{Kennicutt:98}. Additionally, estimates of the dust content based on the UV spectral slope suffer from the degeneracy between a galaxy's dust content, its metallicity and the age of its stellar population, where more dust, more metals or older stars will similarly cause a redder spectral slope. The availability of a spectroscopic measurement of the metallicity, combined with multi-wavelength SED fitting with coverage of the age-sensitive 4000\AA\ break addresses both problems by fully characterizing the galaxy's stellar population, including reliable independent estimates of the dust content, metallicity and age. For these reasons, we will only use the SFR reported by the SED fit in our final comparison of SFR indicators in \S\ref{sec:resultssfr}. However, since multi-wavelength photometry extending redward of the 4000\AA\ break is not straightforward to obtain and metallicity measurements become increasingly uncertain for galaxies at $z>1$, we also quantify how reliably we can estimate the dust extinction from the UV spectral slope in \S\ref{sec:resultssed}.

\subsubsection{Star formation from infrared emission.}
\label{sec:sfrir}
An independent estimate of the dust extinction can be obtained from a galaxy's infrared luminosity. Dust grains will re-radiate the absorbed UV emission at infrared wavelengths, such that a galaxy's combined UV and IR luminosity is a good proxy for the total bolometric output from its newly formed stars. One caveat is the contribution to the infrared emission from dust heated by old stars. We can reasonably assume this contribution to be negligible for this sample of galaxies, where the observed SEDs are best fitted by young stellar populations and we found strict limits on the presence of older stars (see \S\ref{sec:methsed}). 
We use the SFR conversion from \cite{Bell:05}\footnotemark[6], which is derived for a $\sim$100\,Myr old stellar population with constant star formation. 
\begin{equation}
\frac{SFR}{\mathrm{M}_\odot \mathrm{yr}^{-1}} ={2.24 \times 10^{-44} (L_{IR}+1.9L_{1600})}{\mathrm{erg~s}^{-1}}
\label{eq:sfruv+ir}
\end{equation}
$L_{IR}$ is the total infrared luminsosity (5-1000\,$\mu$m) and $L_{1600}=\nu l_{\nu,1600}$. The factor 1.9 accounts for the spectral slope of the stellar population, such that $1.9 L_{1600}$ includes all the UV emission.
\footnotetext[6]{\cite{Bell:05} use the Kroupa IMF, we convert to the Chabrier IMF with SFR$_{\mathrm{Chab}}$ = 0.88 SFR$_{\mathrm{Kroupa}}$.}
\npar
$L_{IR}$ is preferably evaluated from sufficient infrared and submillimeter data to capture the bulk of the bolometric energy output of the infrared SED. Unfortunately, data at these longer wavelengths is hard to obtain, especially for sources at higher redshift, due to confusion noise and sensitivity limitations. For this sample, 24\,$\mu$m data exist for RCSGA0327, cB58 and SGAS1527. cB58 has additionally been detected at 70\,$\mu$m \citep{Siana:08}, 850\,$\mu$m \citep{vanderwerf:01} and 1200\,$\mu$m \citep{Baker:01}. Much work has been done to calibrate $L_{IR}$ from observed 24\,$\mu$m photometry, which traces the mid-IR region (6-12\,$\mu$m) for galaxies at $z\sim2$. This wavelength range is dominated by polycyclic aromatic hydrocarbon (PAH) features, which are stochastically heated by single UV photons. 
Recent studies have shown that the physical conditions in star-forming galaxies at $z\sim2$ are similar to those in local star-forming galaxies of significantly lower infrared luminosity \citep{Papovich:07,Rigby:08,Rujopakarn:11a}, such that when local infrared SEDs are used to extrapolate the observed 24\,$\mu$m emission of $z\sim2$ galaxies, $L_{IR}$ will be overpredicted by factors of 5-10. This is commonly referred to as the mid-IR excess and has been confirmed by recent Herschel results \citep{Elbaz:10, Nordon:10}. We avoid this bias by using a new prescription for $L_{IR}$, based on the local IR SED templates from \cite{Rieke:09}, adapted to $z\sim2$ by accounting for the changing physical conditions of higher redshift star-forming regions \citep{Rujopakarn:11b}. This prescription has a systematic uncertainty of 0.1~dex.

\subsubsection{Star formation from nebular emission lines.}
\label{sec:sfrlines}
The luminosity of the H$\alpha$ recombination line is directly coupled to the incident number of Lyman continuum photons produced by young stars, and hence is proportional to the SFR. 
Star formation rates derived from dust corrected H$\alpha$ emission are generally seen as robust and often used as a comparison and/or calibrator for other indicators (e.g. Gilbank et al. 2010, Reddy et al. 2010). When the H$\alpha$ line is redshifted out of the optical window, the [O~II]~$\lambda$3727 emission line is commonly used as an indicator of star formation. The luminosity of forbidden lines is not directly coupled to the number of ionizing photons, since the excitation depends on the abundance and ionization state of the gas. \cite{Gilbank:10} derived a mass-dependent correction for these effects by calibrating [O~II] SFRs against dust-corrected H$\alpha$ SFRs locally. 

We use the conversions from $L_{H\alpha}$ and $L_{[O~II]}$ to SFR from \cite{Kennicutt:98}\footnotemark[7]. 
The reddening of the nebular gas $E(B-V)_g$ as derived in \S\ref{sec:methav} is used to correct for extinction. For [O~II], we also present values of the corrected [O~II] SFR following \cite{Gilbank:10}\footnotemark[8]. The size of this correction is relatively small over the mass range of the galaxies.

\footnotetext[7]{\cite{Kennicutt:98} uses the Salpeter IMF, we convert to the Chabrier IMF with SFR$_{\mathrm{Salp}}$ = 1.7 SFR$_{\mathrm{Chab}}$.}
\footnotetext[8]{\cite{Gilbank:10} use the Kroupa IMF, we convert to the Chabrier IMF with SFR$_{\mathrm{Chab}}$ = 0.88 SFR$_{\mathrm{Kroupa}}$.}

\section{Results}
\label{sec:results}
\subsection{Physical conditions from rest-frame optical spectroscopy}
\label{sec:resultsspec}
Analysis of the nebular emission lines from rest-frame optical spectroscopy may allow constraints to be placed on extinction, redshift, velocity width, electron density, ionization parameter and metallicity of galaxies. The results for SGAS1527 and SGAS1226 from the spectra presented in \S\ref{sec:spec} are summarized below. 

\begin{itemize}
\item{\textit{Extinction}}\\
We detect two Balmer lines for SGAS1527, H$\alpha$ and H$\beta$, in two different filters. Unfortunately, because the N7 observation lacks a contemporary telluric standard, we do not trust the relative fluxing between H$\alpha$ and H$\beta$ to measure the reddening. H$\gamma$ is not covered, and H$\delta$ is lost to a skyline. For SGAS1226 we detect only one Balmer line, H$\beta$. H$\gamma$ and H$\delta$ were covered by the Nirspec-5b filter, but the lines were not detected. H$\alpha$ is not accessible from the ground. Thus, we are not able to reliably measure an extinction from the Balmer lines for either galaxy.  

\item{\textit{Redshift}}\\
We fit the nebular redshift of SGAS1527 using  the following emission lines: [Ne~III]~$\lambda$3869, H$\beta$, [O~III]~$\lambda$4959,5007 and H$\alpha$. The derived redshift is $z=2.76195\pm0.0002$. The nebular redshift of SGAS1226 is derived from the emission lines of H$\beta$ and [O~III]~$\lambda$4959,5007 and yields $z=2.9257\pm0.0004$. Both redshifts are slightly higher than those reported in \cite{Koester:10}, which were based on absorption features. The presence of galaxy outflows of $\sim 200$~km/s rest-frame can explain these offsets. The emission line redshifts measured here should serve as the systemic redshifts, since one expects the H~II regions to be at rest with respect to the stars.

\item{\textit{Velocity Width}}\\
We measure velocity widths of bright lines, and compare them to the instrumental resolution as measured from Argon lamp exposures (see \cite{Rigby:11} for more details). Fits to the [O~III]~$\lambda$4959,5007 and H$\alpha$ lines of SGAS1527 result in $\sigma=47\pm4$~km~s$^{-1}$. For SGAS1226, $\sigma=100\pm20$~km~s$^{-1}$ is derived from the [O~III]~$\lambda$4959,5007 lines. H$\beta$ and the [O~II]~$\lambda$3727 doublet do not inform this analysis, since H$\beta$ is too weak for a useful measurement of its linewidth, and since the [O~II]~$\lambda$3727 doublet is not sufficiently resolved for a non-degenerate fit. These values are consistent with the broad range of $\sigma\sim40-200$~km~s$^{-1}$ found at $z\sim2-3$ \citep{Erb:06b}.

\item{\textit{Electron density}}\\
\cite{Rigby:11} derived an electron density from a two-component fit to the flux ratio of the [O~II]~$\lambda$3727 doublet of RCSGA0327. We have attempted this same procedure for SGAS1527 and SGAS1226, but the lower signal-to-noise of the spectra leads to ambiguous fitting results. In SGAS1527, the line profile appears to have two components, but fitting it as such returns unphysically narrow linewidths - half the instrumental resolution at this wavelength. The source  of the problem may be the sky line that falls on the bluer line of the doublet. As a result, we cannot simultaneously determine both the doublet ratio and the velocity dispersion $\sigma$ in the [O~II]~$\lambda$3727 doublet. Instead, we vary $\sigma$ within the range permitted by the fit to the [O~III]~$\lambda$4959,5007 lines, fitting a doublet ratio for each value. This results in a flux ratio f(3727/3729) $=1.03\pm0.2$. Using the IRAF task stsdas.analysis.nebular.temden, our measurement corresponds to an electron density $n_e = 400^{+260}_{-230}$~cm$^{-3}$ at $T_e=10^4$~K. Following a similar procedure for SGAS1226, we find a flux ratio of f(3727/3729) $=0.59\pm0.1$. This corresponds to the low-density limit for  $T_e = 10^4$~K. 

\cite{Rigby:11} found a tight constraint of $n_e = 235^{+28}_{-26}$~cm$^{-3}$ (at $T_e=10^4$~K) for RCSGA0327. They summarize constraints from the literature for the Clone, the Cosmic Horseshoe and J0900+2234 (see \cite{Rigby:11} for references) as a wide range of $n_e = 600-5000$~cm$^{-3}$, with large uncertainties for individual measurements. These constraints are remarkably higher than the low electron densities found for RCSGA0327, SGAS1527 and SGAS1226, which lie closer to the typical densities of local H~II regions, $n_e\sim100$\,cm$^{-3}$ (e.g. Zaritsky et al. 1994). 

The inverse relation between electron density and size of the H~II regions found locally \citep{KimKoo:01} predicts that local H~II regions with $n_e\sim100$\,cm$^{-3}$ should have diameters of $\sim 8$\,pc. If this relation holds at $z\sim2$, $n_e\sim300$\,cm$^{-3}$ as measured on average for RCSGA0327, SGAS1527 and SGAS1226 corresponds to star-forming regions of $\sim 3$\,pc, while $n_e\sim1000$\,cm$^{-3}$ as an average for the Clone, the Cosmic Horseshoe and J0900+2234 implies H~II regions as small as 1\,pc. However, the uncertainties on the electron density measurements of these latter sources are large; rest-frame optical spectra of similar quality to what we present in this paper and have published for RCSGA0327 \citep{Rigby:11} is needed for a larger sample of individual star-forming galaxies at $z\sim2$.

\item{\textit{Ionization parameter}}\\
Figure~1 of \cite{Kewley:02} illustrates the use of the flux ratio of [O~III]~$\lambda$5007 to [O~II]~$\lambda$3727 as a diagnostic of the ionization parameter. Assuming an oxygen abundance of 20-40$\%$ solar (on the Asplund system), Equation~12 of \cite{Kewley:02} \footnotemark[9] yields an ionization parameter of $\log U = -2.9$ to $-2.7$ for SGAS1527 and $\log U = -2.8$ to $-2.9$ for SGAS1226. This assumes no extinction; an extinction correction lowers the inferred $\log U$. These results increase the number of lensed galaxies for which this diagnostic is measured from four to six. As noted in \cite{Rigby:11}, they all have very similar ionization paremeters of $\log U = -2.9$ to $-2.7$. This is roughly twice as high as $-3.2 < \log U < -2.9$ measured for lower luminosity local galaxies (Moustakas et al. 2006, 2010).
\footnotetext[9]{\cite{Kewley:02} use the abundances from \cite{Anders:89}; we convert to \cite{Asplund:09}.}

\item{\textit{Oxygen abundance}}\\
Unfortunately, the ``gold standard'' diagnostic of oxygen abundance, [O~III]~$\lambda$4363, is unavailable in either galaxy, due to limited wavelength coverage in SGAS1527 and skyline contamination in SGAS1226. The ratio of [N~II]~$\lambda$6583 to H$\alpha$, known as the N2 index, is a commonly-used, reddening-free abundance indicator. Unfortunately, the redshift of SGAS1226 is too high for these lines to be visible in the near-IR windows. The lines are covered in SGAS1527, though [N~II] is not detected, limiting us to an upper limit on the oxygen abundance. We simultaneously fit H$\alpha$ and the two [N~II] lines, setting a common linewidth for all three lines that is allowed to vary within the range measured above. The $3\sigma$ upper limit is $\log$~([N~II]~$\lambda$6583/H$\alpha$) $< -0.65$, which corresponds to an oxygen abundance of $12 + \log(O/H) < 8.5$ and a metallicity of $<70\%$ of solar on the \cite{Asplund:09} scale. The $1\sigma$ upper limit on the oxygen abundance is $12 + \log(O/H) < 8.3$, which translates to $<40\%$ of solar. 

The Ne3O2 index, the ratio of [Ne~III]~$\lambda$3869 to [O~II]~$\lambda$3727 is another accessible, reddening-free abundance indicator. Unfortunatley the [Ne~III] line is lost to a skyline in the spectrum of SGAS1226. For SGAS1527, the measured ratio of $\log$~([Ne~III]~$\lambda$3869/[O~II]~$\lambda$3727)$=-0.42^{+0.12}_{-0.17}$ yields an abundance of $12 + \log(O/H) = 7.6 \pm 0.2$ via the calibration of \cite{Shi:07}. This corresponds to a metallicity of $5$-$12\%$ of solar on the \cite{Asplund:09} scale. Unfortunately, as Figure~1 of \cite{Shi:07} makes clear, the scatter in this calibration is large, 0.3 dex overall and worse at low ratios of [Ne~III]/[O~II]. This diagnostic does therefore not prove useful in this regime.
\end{itemize}

\subsection{Stellar population parameters}
\label{sec:resultssed}
\begin{figure}[h]
\centering
\includegraphics[width=9cm]{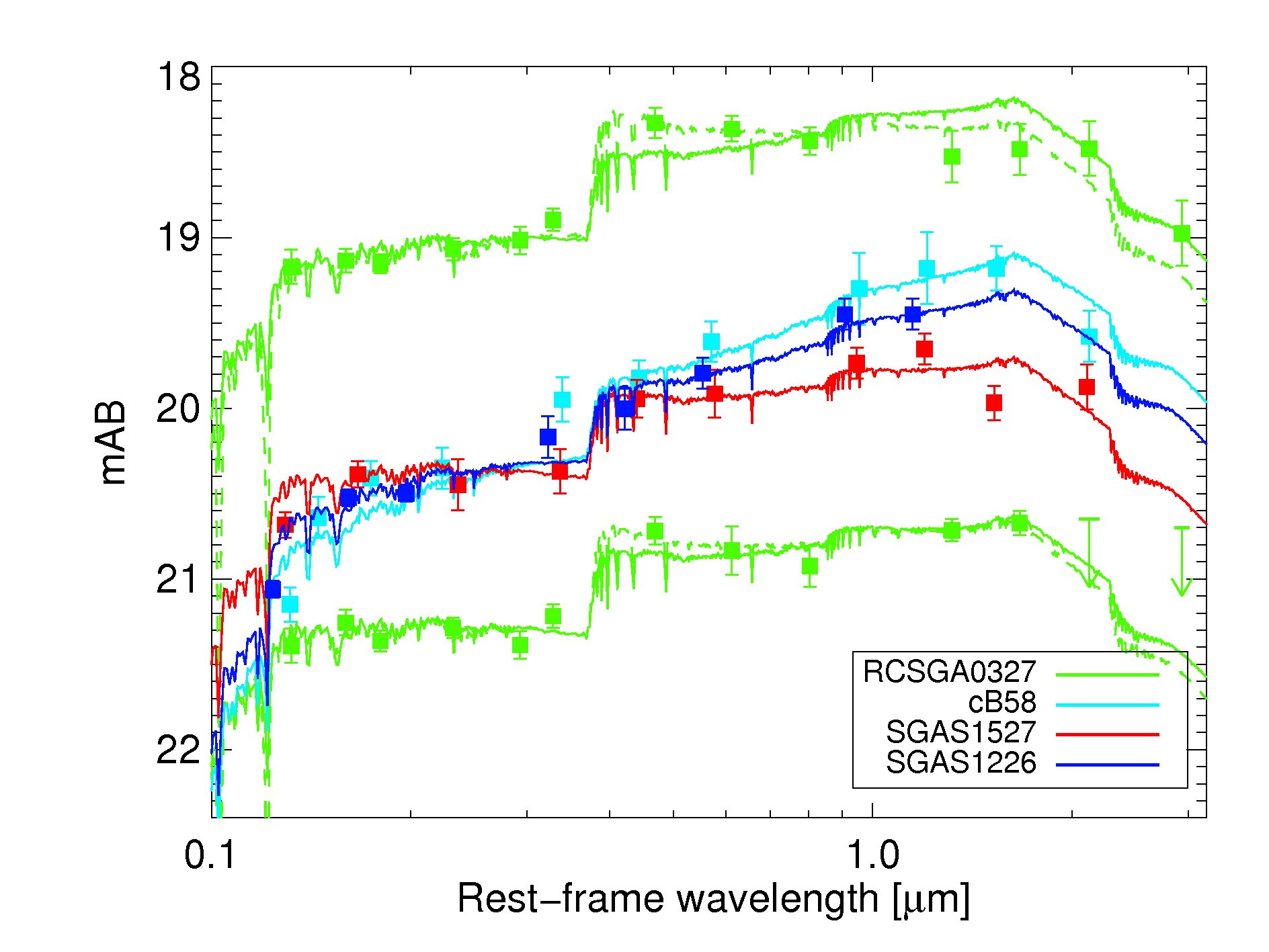}
\caption{Best fit stellar energy distributions for the arc \textit{(top)} and counter-image \textit{(bottom)} of RCSGA0327, cB58, SGAS1527 and SGAS1226 derived from age-restricted CSF models. The dashed green lines show the improved SED fits for both the arc and counter-image of RCSGA0327 when using an exponentially declining SFH with e-folding time $\tau$=10-50\,Myr. Photometry datapoints are overplotted with $1\sigma$ errorbars. \label{fig:seds}}
\end{figure}

The stellar populations of the lensed galaxies are constrained by the rest-frame UV to near-IR spectral energy distributions. Figure~\ref{fig:seds} shows the best-fit SEDs for the arc and counter-image of RCSGA0327, cB58, SGAS1527 and SGAS1226 at rest-frame wavelengths. We have investigated whether the small discrepancy between the photometry and the best-fit SED model that can be seen for RCSGA0327, cB58 and SGAS1226 around the 4000\AA\ break can be significantly reduced by allowing exponentially declining SFHs. We find that this is not the case for cB58 and SGAS1226. For these galaxies the discrepancy is due to a tension between our minimum age constraint of 70\,Myr (well motivated in \S\ref{sec:methsed}) and the lack of a significant 4000\AA\ break in the observed photometry. For RCSGA0327, SED models with an e-folding time of $\tau=10-50$\,Myr provide improved fits to the photometry, shown by the dashed green lines in Figure~\ref{fig:seds}. The current SFR derived from the SED fit depends greatly on the assumed star formation history. For both the arc and the counter-image of RCSGA0327, exponentially declining SFHs with $\tau=10-50$\,Myr produce SFRs $<20$~M$_\odot$\,yr$^{-1}$, much smaller than indicated by the other SFR measurements (see \S\ref{sec:resultssfr}). Recently, \cite{Wuyts:11} have similarly reported the need for a minimum e-folding time of $\log(\tau)=9.5$ (for CB07models) to avoid SFRs that are underestimated compared to measurements derived from the bolometric luminosity for galaxies out to $z\sim3$. For the relatively young ages of the lensed galaxies in this sample, such long e-folding times are equivalent to the assumption of constant star formation.

Table~\ref{tab:sed} reports the stellar population parameters derived from the SED fit. The confidence intervals are estimated from 1000 mock realizations of the observed SEDs which are consistent with the photometric uncertainties, and by varying the age restriction between 50 and 100\,Myr. The age $t$ and reddening factor $E(B-V)$ depend solely on the galaxy's colors; the stellar mass and star formation rate have an additional dependence on the lensing magnification. Based on WFC3/HST images, \cite{Sharon:11} (in preparation) report a magnification factor of 3.0$\pm$0.2 for the counter-image of RCSGA0327 and an average magnification of 25.1$\pm$3.2 across the giant arc. \cite{Koester:10} report magnification limits of $<15$ for SGAS1527 and $>40$ for SGAS1226. cB58 is magnified by a factor of 30, with an estimated uncertainty of 30\% \citep{Seitz:98}. The magnification uncertainties are included in the reported uncertainties for the stellar mass and SFR.

\begin{deluxetable*}{lcccc} 
\tablecolumns{6} 
\tablecaption{Stellar population parameters. \label{tab:sed}} 
\tablehead{ \multicolumn{1}{c}{source} &
	    \multicolumn{1}{c}{Age}  & 
            \multicolumn{1}{c}{$E(B-V)_s$}  & 
	    \multicolumn{1}{c}{$\log(\mathrm{M}_*/\mathrm{M}_\odot)$}         &
	    \multicolumn{1}{c}{$SFR$}  \\
	    \multicolumn{1}{c}{} &
	    \multicolumn{1}{c}{(Myr)}  & 
            \multicolumn{1}{c}{}  & 
	    \multicolumn{1}{c}{}         &
	    \multicolumn{1}{c}{(M$_\odot$\,yr$^{-1}$)}} 
\startdata 
RCSGA0327-A  & 128$\pm$13  & 0.15$\pm$0.035 & 9.83$\pm$0.06  & 71$\pm$12  \\
RCSGA0327-C  & 143$\pm$42  & 0.10$\pm$0.04  & 9.76$\pm$0.05  & 52$\pm$15  \\ 
cB58         & 72$\pm$18   & 0.25$\pm$0.055 & 9.69$\pm$0.14  & 82$\pm$27  \\
SGAS1527     & 126$\pm$31  & 0.15$\pm$0.05  & $>9.82\pm0.04$ & $>71\pm16$ \\ 
SGAS1226     & 86$\pm$41   & 0.20$\pm$0.05  & $<9.54\pm0.07$ & $<48\pm15$ \\ 
\enddata
\end{deluxetable*}

The stellar populations of all four lensed galaxies are remarkably similar and can be characterized on average with an age $\sim$ 110\,Myr, reddening $E(B-V)_s \sim 0.17$, SFR $\sim$ 65\,M$_\odot$~yr$^{-1}$ and stellar mass $\sim 5.5 \times 10^9$\,M$_\odot$. Compared to representative samples of UV-selected star-forming galaxies at $z\sim2-3$ \citep{Shapley:01,Erb:06b,Mannucci:09,Reddy:10}, these lensed galaxies represent young, starbursting systems with low stellar masses, suggesting they have only recently begun building their stellar population. 
The same conclusion is reached from a comparison of the UV luminosity and stellar mass to the characteristic luminosity and stellar mass of comparable galaxies at $z\sim2-3$. Using the characteristic luminosity from \cite{Oesch:10} at $z=1.5-2.0$ for a comparison to RCSGA0327 and the results from \cite{Reddy:09} at $z=2.7-3.4$ for a comparison to cB58, SGAS1527 and SGAS1226, we find an average brightness of $\sim 2.6$~L$_*$. The stellar mass of the sample lies at $\sim 0.08$~M$_{star}^*$ when comparing to the characteristic stellar mass found by \cite{Marchesini:09} at $z\sim2-3$.
\npar
We now revisit the issue raised in \S\ref{sec:methsfr} of how reliably one can estimate the dust extinction $E(B-V)_s$ from the UV spectral slope alone. This is relevant for galaxies at $z>1$, where the photometric and spectroscopic data required to fully characterize the galaxy's stellar population through SED fitting becomes hard to obtain. We use the UV color of the galaxies, derived from the filters that best match the rest-frame 1600\AA-2300\AA\ window, to measure the UV spectral slope $\beta$ ($f_\lambda \sim \lambda^\beta$). The reddening $E(B-V)_s$ derived from $\beta$ and the relation established by \cite{Meurer:99} for local starburst galaxies is consistently a factor of 1.25-2.0 lower than the value reported by the SED fit. This underestimates the dust-corrected SFR of the galaxies by factors of 2-3.5. 

\begin{figure}[h]
\centering
\includegraphics[width=9cm]{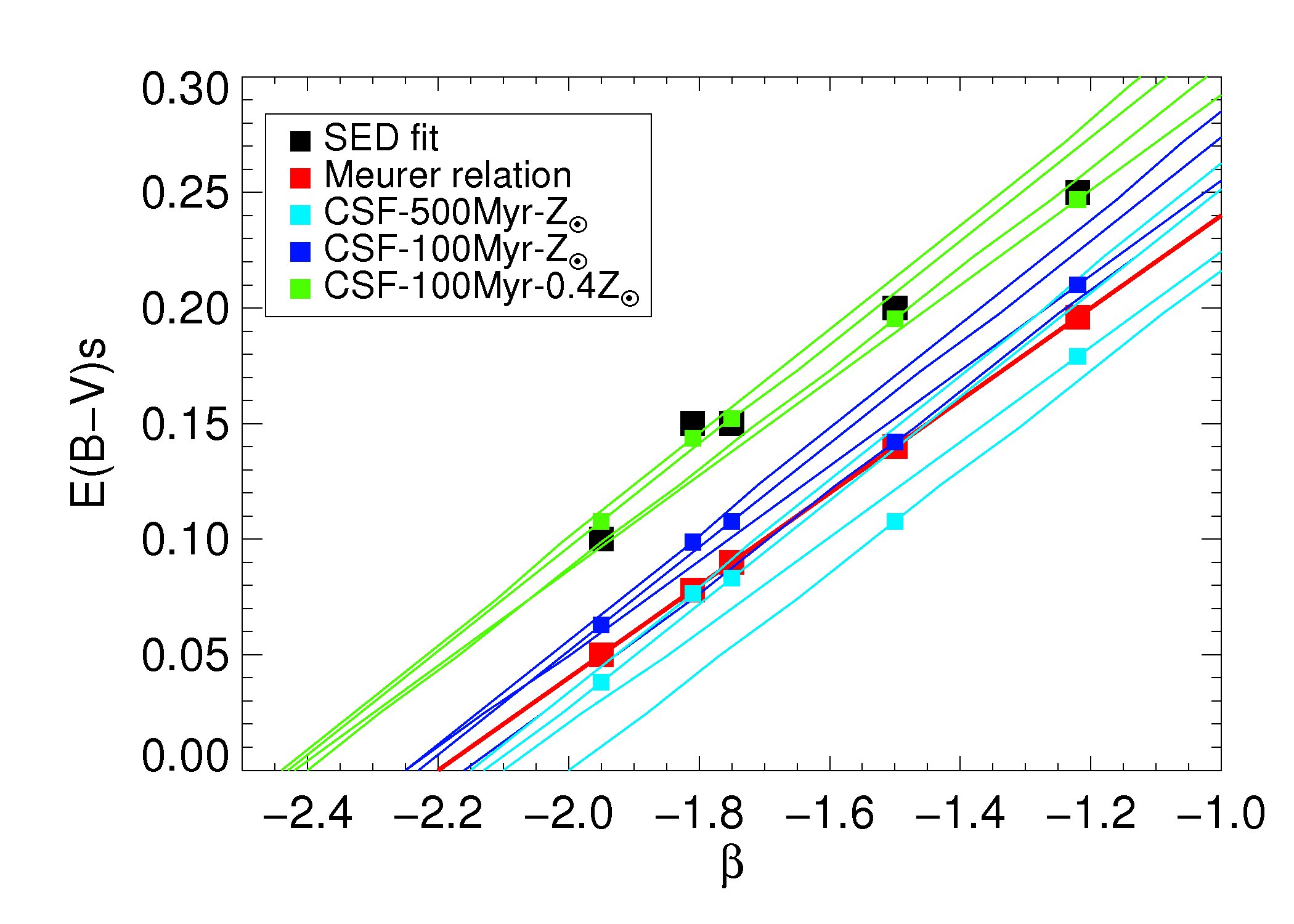}
\caption{The relation between the UV spectral slope $\beta$ and the reddening $E(B-V)_s$. The local relation established by \cite{Meurer:99} is shown in red. The black datapoints show the galaxies' best-fit $E(B-V)_s$ from the SED fit. The relations we derived for the SED models are shown in cyan, blue and green. The 100\,Myr, 0.4~Z$_\odot$, CSF model is representative of the average stellar population of the galaxies and agrees with the results from the SED fit. \label{fig:beta_to_ext}}
\end{figure}

We investigate the origin of this discrepancy by establishing the relation between $\beta$ and $E(B-V)_s$ for representative CB07 models. For a range of values $E(B-V)_s=[0,0.3]$ we redden each SED model with the Calzetti dust extinction law and measure the UV spectral slope based on the same filters available to us for each of the four galaxies in the sample (specifically $B$ and $r$ for RCSGA0327, $V$ and $I$ for cB58, $r$ and $z$ for SGAS1527 and $r$ and $i$ for SGAS1226). Since $\beta$ depends somehat on the exact wavelength window used, this results in four slightly different relations between $\beta$ and $E(B-V)_s$ for each SED model. Figure~\ref{fig:beta_to_ext} shows the Meurer relation in red, offset from the $E(B-V)_s$ values reported from the SED fit (black datapoints). The relations between $\beta$ and $E(B-V)_s$ for a solar metallicity, 500~Myr old CSF model and a 100\,Myr old CSF model with a metallicity of Z$_\odot$ and 0.4~Z$_\odot$ are shown in cyan, blue and green respectively. The 100\,Myr, 0.4~Z$_\odot$, CSF model is representative of the average stellar population of the galaxies, and can be seen to agree with the best-fit values for $E(B-V)_s$ from the SED fit. The lower metallicity accounts for much of the discrepancy with the Meurer relation: a 0.4~Z$_\odot$ SED model is intrinsically less red than a solar metallicity SED model and thus requires a larger dust correction to match the same UV spectral slope. In general, this highlights the danger of blindly adopting the Meurer relation to dust-correct the UV continuum emission of high redshift galaxies, without additional information on the age and metallicity of the stellar populations. 

\subsection{Reddening}
\label{sec:resultsav}
\begin{deluxetable}{lccc}[h] 
\tablecolumns{4} 
\tablecaption{Extinction estimates. \label{tab:av}} 
\tablehead{ \multicolumn{1}{c}{source} &
 	    \multicolumn{1}{c}{$E(B-V)_s$} &
	    \multicolumn{1}{c}{$E(B-V)_g^{Balmer}$} &
 	    \multicolumn{1}{c}{$E(B-V)_g^{M_*}$}}
\startdata 
RCSGA0327-A  & 0.15$\pm$0.035 & 0.20$\pm$0.19 & 0.31$\pm$0.18 \\
RCSGA0327-C  & 0.10$\pm$0.04 & \nodata	     & 0.29$\pm$0.17 \\
cB58         & 0.25$\pm$0.055 & 0.11$\pm$0.10 & 0.20$\pm$0.12 \\
SGAS1527     & 0.15$\pm$0.05 & \nodata	     & 0.24$\pm$0.14 \\
SGAS1226     & 0.20$\pm$0.05 & \nodata	     & 0.15$\pm$0.09 \\
\enddata
\footnotetext[]{Column 1: Reddening of the stellar light from the SED fit}
\footnotetext[]{Column 2: Reddening of the nebular gas from the Balmer decrement}

\footnotetext[]{Column 3: Reddening of the nebular gas based on the stellar mass as derived by \cite{Gilbank:10}}
\end{deluxetable}
We report the reddening of the stellar light, $E(B-V)_s$ and the reddening of the ionized gas, $E(B-V)_g$ in Table~\ref{tab:av}. These measures of reddening are important to reliably correct the SFR indicators for dust extinction. $E(B-V)_s$ is taken from the SED fit, $E(B-V)_g$ is estimated from the stellar mass following the local G10 relation (Gilbank et al. 2010, see \S\ref{sec:methav}). For the giant arc of RCSGA0327 and for cB58, a measurement of the gas reddening can also be made from the Balmer decrement \citep{Rigby:11,Teplitz:00} and the independent estimates of $E(B-V)_g$ appear consistent within the $1\sigma$ uncertainties. This suggests that the local G10 relation remains valid at $z\sim2$ and there is mild or no evolution in the dust content of galaxies as a function of mass between these epochs. However, we note that this is only based on two measurements and more work is needed to verify the validity of this local prescription at higher redshift. For RCSGA0327 and cB58, we use the weighted average of both measures of $E(B-V)_g$ to correct the nebular emission line SFR indicators for extinction.

Within the $1\sigma$ uncertainties, these reddening measurements cannot conclude whether or not the nebular emission lines experience larger reddening than the stellar light in this sample of star-forming galaxies.

\subsection{Star Formation Rates}
\label{sec:resultssfr}
We have outlined 4 different SFR indicators in \S\ref{sec:methsfr} based on the SED fit, the total bolometric luminosity and the H$\alpha$ and [O~II]~$\lambda$3727 emission lines. The SFR result from the SED fit is reported in \S\ref{sec:resultssed}, after being corrected for the lensing magnification ($25.1\pm3.2$ for RCSGA0327-A, $3.0\pm0.2$ for RCSGA0327-C, $30\pm9$ for cB58, $<15$ for SGAS1527 and $>40$ for SGAS1226). We summarize the de-lensed UV and IR luminosities of the galaxies (corrected using the same magnification factors) in Table~\ref{tab:lum} and use Equation~\ref{eq:sfruv+ir} to derive the SFR from the bolometric luminosity.

\begin{deluxetable*}{lcccc}[h]
\tablecolumns{5} 
\tablecaption{De-lensed luminosities. \label{tab:lum}} 
\tablehead{ \multicolumn{1}{c}{source} &
	    \multicolumn{1}{c}{$L_{1600}$} &
	    \multicolumn{1}{c}{$L_{IR}$} &
	    \multicolumn{1}{c}{$L_{H\alpha}$} &											
	    \multicolumn{1}{c}{$L_{[O~II]}$} \\ 
 	    \multicolumn{1}{c}{} &
            \multicolumn{1}{c}{$10^{44}$\,erg~s$^{-1}$} &
  	    \multicolumn{1}{c}{$10^{44}$\,erg~s$^{-1}$} &  
	    \multicolumn{1}{c}{$10^{42}$\,erg~s$^{-1}$} &  
	    \multicolumn{1}{c}{$10^{42}$\,erg~s$^{-1}$}} 								
\startdata
RCSGA0327-A & 4.3$\pm$0.6  & 8.4$\pm$3.9                     & 5.4$\pm$1.4\footnotemark[a]    & 3.3$\pm$0.4\footnotemark[a] \\
RCSGA0327-C & 5.1$\pm$0.5  & 2.4$\pm$1.3                     & \nodata                        & \nodata                     \\
cB58        & 2.1$\pm$0.7  & 4.9$\pm$2.2                     & 2.6$\pm$0.8\footnotemark[b]    & 2.6$\pm$0.8\footnotemark[b] \\ 
SGAS1527    & $>5.3\pm0.4$ & $\sim3.7\pm1.7$\footnotemark[c] & $>14\pm8$\footnotemark[d]      & $>4.6\pm0.9$                \\ 
SGAS1226    & $<2.0\pm0.1$ & \nodata                         & $<1.1\pm0.4$\footnotemark[d]   & $<1.7\pm0.3$                \\ 
\enddata
\footnotetext[a]{\cite{Rigby:11} - corrected for an average magnification of $42.2\pm5.5$ as calculated by Sharon et al. (2011, in preparation). These luminosities are derived from the area of the giant arc targeted for spectroscopy, which translates into a small region of the galaxy in the source plane (Sharon et al. 2011, in preparation), and are not valid for the whole galaxy.}
\footnotetext[b]{\cite{Teplitz:00}}
\footnotetext[c]{SGAS1527 was not detected at 24\,$\mu$m, the reported luminosity represents a $3\sigma$ upper limit, corrected for a lower limit on the magnification of $<15$}
\footnotetext[d]{H$\alpha$ is derived from H$\beta$ and the Balmer decrement as derived from the stellar mass through the G10 relation \citep{Gilbank:10}} 
\end{deluxetable*}

The emission line luminosities of SGAS1527 and SGAS1226 are derived from the observed fluxes reported in Table~\ref{tab:fluxes}. As can be seen in Figure~\ref{fig:findercharts}, the NIRSPEC slit covers the whole source for both galaxies, so we can use the total magnifications of $<15$ for SGAS1527 and $>40$ for SGAS1226 to calculate the de-lensed luminosities, without additional aperture corrections. Since no measurement of H$\alpha$ exists for either source (we don't trust the fluxing of SGAS1527 due to a time delay with the observation of the telluric star and for SGAS1226 H$\alpha$ falls outside the observable window due to its high redshift), we use H$\beta$ and the extinction $E(B-V)_g$ derived from the stellar mass following the G10 relation (see \S~\ref{sec:resultsav}) to estimate $L_{H\alpha}$. For cB58, the H$\alpha$ and [O~II]~$\lambda$3727 luminosities published by \cite{Teplitz:00} are used and corrected for a total magnification of $30\pm9$. For RCSGA0327, we use the emission line measurements from \cite{Rigby:11}. Sharon et al. (2011, in preparation) calculate an average magnification of $42.2\pm5.5$ for the area of the arc covered by the NIRSPEC slit. \cite{Rigby:11} recommend using the observed H$\beta$ flux and the Balmer decrement to estimate $L_{H\alpha}$. For all galaxies, de-lensed H$\alpha$ and [O~II]~$\lambda$3727 luminosities are summarized in Table~\ref{tab:lum} and the resulting SFRs are dust-corrected using the Calzetti extinction law and $E(B-V)_g$ from \S~\ref{sec:resultsav}.

\begin{figure}[h]
\centering
\includegraphics[width=9cm]{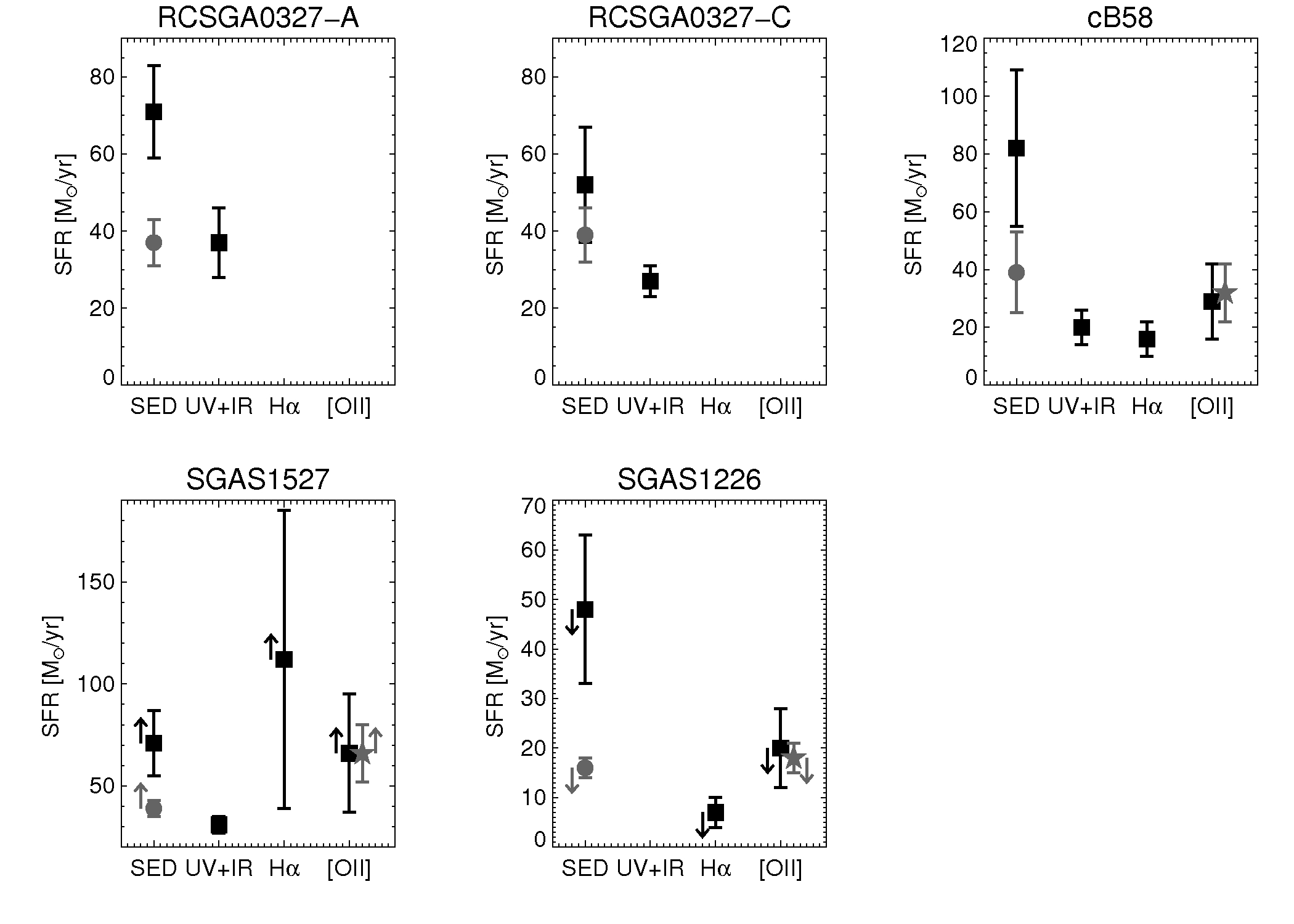}
\caption{Comparison of SFR indicators. The gray stars correspond to the mass-dependent G10 correction for $SFR_{[O~II]}$ from \cite{Gilbank:10}. The SFR derived from the SED fit is overall larger than the other indicators, which suggests that the Calzetti extinction law overpredicts the dust correction for the UV luminosity. The results from an SED fit with the steeper SMC extinction law are shown as gray circles. \label{fig:sfrcomp}}
\end{figure}

The different SFR indicators are compiled in Table~\ref{tab:sfr} and Figure~\ref{fig:sfrcomp} for easy comparison. For SGAS1527 and SGAS1226, all SFR measurements are upper/lower limits due to the poorly constrained magnification; however on a relative scale, these limits can still be used for a comparison of the different indicators.

\begin{deluxetable*}{lccccc}[h]
\tablecolumns{6} 
\tablecaption{Star Formation Rates. \label{tab:sfr}} 
\tablehead{ \multicolumn{1}{c}{source} &
	    \multicolumn{1}{c}{$SFR_{SED}$} &
	    \multicolumn{1}{c}{$SFR_{UV+IR}$} &
	    \multicolumn{1}{c}{$SFR_{H\alpha}$} &											
	    \multicolumn{1}{c}{$SFR_{[O~II]}$} &
            \multicolumn{1}{c}{$SFR_{SED-SMC}$}}   								
\startdata 
RCSGA0327-A & 71$\pm$12  & 37$\pm$9                     & 44$\pm$17\footnotemark[a]   & 48$\pm$21\footnotemark[a] & 37$\pm$6  \\
RCSGA0327-C & 52$\pm$15  & 27$\pm$4                     & \nodata                     & \nodata                   & 39$\pm$7  \\
cB58        & 82$\pm$27  & 20$\pm$6                     & 16$\pm$6                    & 29$\pm$13                 & 38$\pm$14 \\
SGAS1527    & $>71\pm16$ & $\sim31\pm4$\footnotemark[b] & $>112\pm73$                 & $>66\pm29$                & 39$\pm$4  \\
SGAS1226    & $<48\pm15$ & \nodata                      & $<7\pm3$                    & $<20\pm8$                 & 16$\pm$2  \\
\enddata
\footnotetext[a]{These SFRs are derived from the area of the giant arc targeted for spectroscopy, which translates into a small region of the galaxy in the source plane (Sharon et al. 2011, in preparation), and are not valid for the whole galaxy.}
\footnotetext[b]{SGAS1527 was not detected at 24\,$\mu$m, the reported luminosity represents a $3\sigma$ upper limit, corrected for a lower limit on the magnification of $<15$}
\end{deluxetable*}
\npar
The complex lensing signature of RCSGA0327 does not allow a direct comparison of the SFRs between the giant arc and the counter-image. The lens model tells us that the counter-image is a more or less uniformly magnified image of the source plane galaxy. The giant arc on the other hand consists of three merged images of the source galaxy and depending on the location with respect to the lensing caustic, different regions within the source plane galaxy receive very different magnifications (see Sharon et al. 2011, in preparation, for more details). The de-lensed luminosities and derived SFRs of the giant arc are therefore not representative of the source galaxy, and should not be compared directly to the results from the counter-image. Additionally, Sharon et al. 2011 (in preparation) find that the area of the arc observed spectroscopically translates into a very small region of the galaxy in the source plane, which consists of a few highly magnified individual star-forming complexes. The SFRs measured for this area from the H$\alpha$ and [O~II]~$\lambda$3727 emission lines are not valid for the whole galaxy and we do not include them in our overall comparison of SFR indicators. A future paper will constrain the H$\alpha$ luminosity of the source galaxy from rest-frame optical spectroscopy of the counter-image.

\npar
Figure~\ref{fig:sfrcomp} shows a general trend where the SFR derived from the SED fit is overall larger than the other indicators (except for SGAS1527). This suggests that the SED fit, and thus the dust-corrected UV continuum emission, overpredicts the star formation rate. A similar result was found for young LBGs (age $<$ 100~Myr) at $z\sim2$ \citep{Reddy:10}, and for cB58 and the Cosmic Eye \citep{Siana:09}. These authors have suggested that the Calzetti extinction law is not a good representation of the dust geometry in young LBGs at $z\sim2$ and will significantly overpredict the dust extinction. 

\begin{figure}[h]
\centering
\includegraphics[width=9cm]{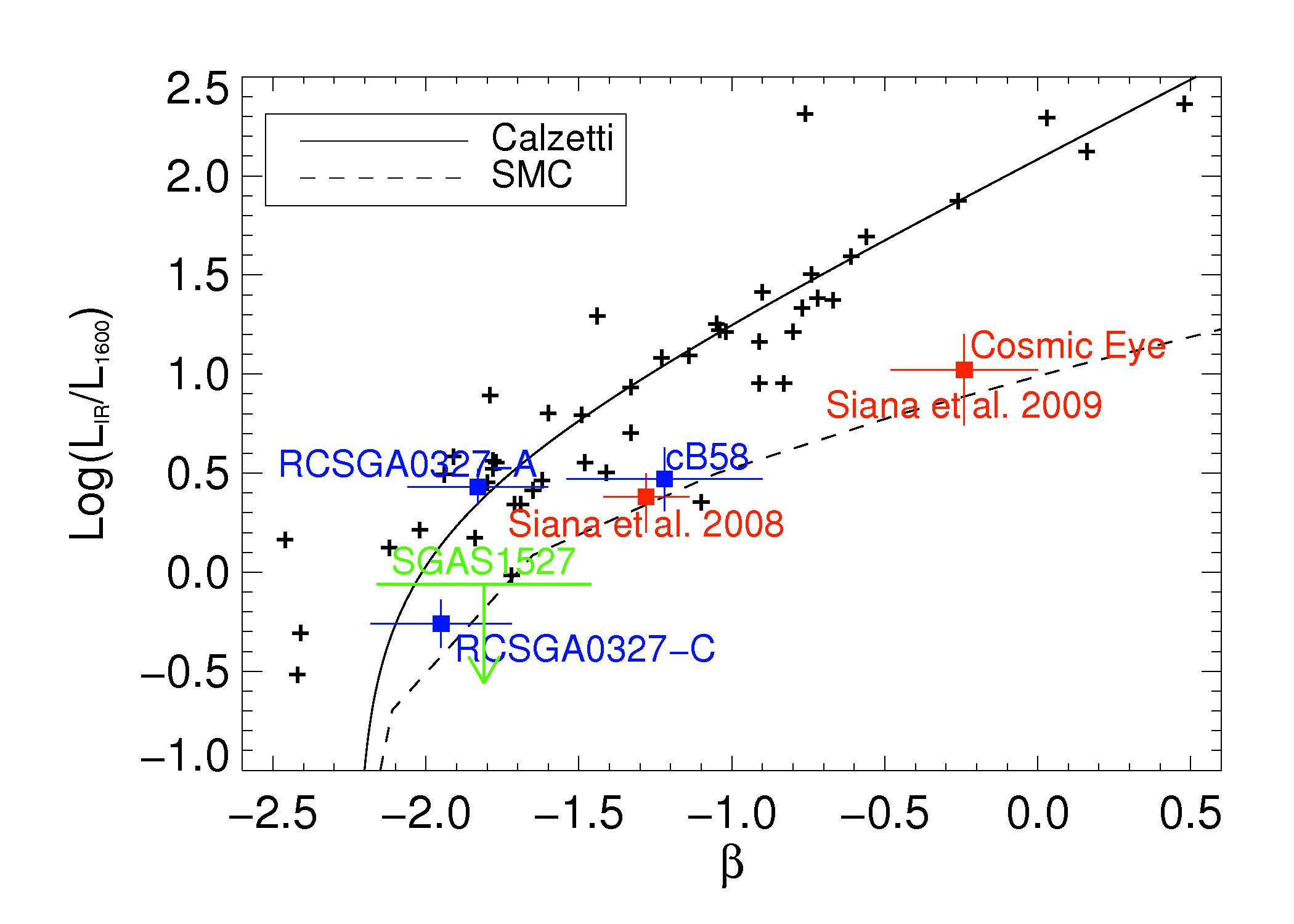}
\vspace{0.05in}
\caption{Dust attenuation, parameterized as the ratio between IR and UV luminosity, versus rest-frame UV spectral slope $\beta$. The measurements of local starburst galaxies are plotted as little black crosses. The best-fit relation corresponds to the Calzetti dust extinction law \citep{Meurer:99}. The steeper SMC extinction law is also shown. The red datapoints for cB58 and the Cosmic Eye are taken from \cite{Siana:08} and \cite{Siana:09} respectively. The relatively blue spectral slopes of RCSGA0327 and SGAS1527 have limited leverage to distinguish between the different extinction laws. \label{fig:meurer}}
\end{figure}

The influence of the shape of the extinction law on the dust extinction is commonly illustrated with the Meurer-plot \citep{Meurer:99}, plotting the UV spectral slope $\beta$ ($f_\lambda \sim \lambda^\beta$) versus the ratio of IR to UV luminosity, which serves as a parametrization of the dust extinction\footnotemark[11]. 
\footnotetext[11]{\cite{Meurer:99} define this relation for $L_{FIR}$, integrated from 40 to 120\,$\micron$, where $L_{IR}=1.75\times L_{FIR}$ based on the calibration by \cite{Calzetti:00}.}
On this plot, a steeper extinction law results in comparatively more absorption of photons at blue wavelenghts and thus a redder UV spectral slope $\beta$ for the same extinction. For local starbursts, plotted as little black crosses in Figure~\ref{fig:meurer}, the Calzetti extinction law is the best fit \citep{Meurer:99}. \cite{Reddy:10} have shown that at $z\sim2$, the Calzetti law remains valid for star-forming galaxies with ages $>100$\,Myr. However, the younger LBGs in their sample, as well as cB58 and the Cosmic Eye, all tend to lie below the Calzetti curve, showing less extinction than would be derived from the UV spectral slope. For these sources, the steeper extinction law derived for the Small Magellanic Cloud (SMC, Prevot et al. 1984) seems to be a better fit. 
We can add datapoints for the arc and counter-image of RCSGA0327, for SGAS1527 and for our re-analysis of cB58 to this plot. 
Due to the relatively blue spectral slopes of RCSGA0327 and SGAS1527, these galaxies have limited leverage to distinguish between the different extinction laws. 

We have repeated the SED fitting with the steeper SMC extinction law. With the exception of cB58, the best-fit SED can not be distinguished from the result obtained with the Calzetti law in terms of the reduced $\chi^2$ of the fit. The improved fit for cB58 was also reported by \cite{Siana:08}. The derived SFRs are lower by a factor of 1.3-2 compared to the Calzetti extinction law and therefore in closer general agreement to the other SFR indicators (gray circles in Figure~\ref{fig:sfrcomp}). With the SMC law, the stellar mass is consistent with previous results, but the best-fit ages are higher by a factor of $\sim2$ and there is no sub-population of very young extreme starbursts, which makes the age restriction to older than 70\,Myr unnecessary. 

\section{Summary}
This paper presents a uniform analysis of rest-frame UV to near-IR spectral energy distributions and rest-frame optical spectra for a sample of four of the brightest lensed galaxies at $z\sim2$. Given the limitations of current facilities, such extensive data collection and detailed analysis of stellar populations and physical conditions would not be possible for individual galaxies at this redshift without the magnification induced by gravitational lensing. SED fits to the rest-frame UV to near-IR photometry result in very uniform stellar population parameters across the sample, which we can characterize as young, bright, starbursting systems with little dust content and low stellar masses. We derive a strict limit on the presence of an older stellar population. Restricting the age of the stellar population to be larger than the dynamical timescale of 70~Myr during the SED fitting is necessary to avoid extreme young starburst models and to obtain metallicities consistent with those found from the rest-frame optical spectroscopy. When the local Meurer relation \citep{Meurer:99} is used to estimate the dust correction from the UV spectral slope, $E(B-V)_s$ is underestimated by factors of 1.25-2, leading to underestimates of the dust-corrected UV continuum SFR by factors of 2-3.5. This is largely due to the lower metallicity (0.4~Z$_\odot$) of the galaxies. This result cautions the use of the Meurer relation without further characterization of a galaxy's stellar population and motivates efforts to measure metallicities for larger samples of both lensed and field galaxies at $z\sim2$.

We present SFR estimates based on the SED fit, the UV+IR bolometric luminosity and the H$\alpha$ and [O~II]~$\lambda$3727 emission lines. The SED fit with the default Calzetti extinction law reports a SFR estimate that is consistently higher than the other indicators. Similar results were found for cB58 and the Cosmic Eye \citep{Siana:09}. This suggests that the Calzetti extinction law is too flat for young star-forming galaxies at $z\sim2$ and therefore overpredicts the dust extinction. In the diagnostic Meurer plot, the galaxies' spectral slopes are unfortunately too blue to differentiate between the Calzetti law and the steeper SMC extinction law. SED fitting with the SMC law returns SFRs lower by a factor of 1.3-2, bringing this method into general agreement with the other SFR indicators, within the $1\sigma$ uncertainties. These results are a first attempt to compare the common SFR indicators in detail for individual galaxies at $z\sim2$; the comparison would clearly benefit from a larger sample of galaxies. 

\begin{acknowledgments}
We thank the referee for useful comments and suggestions that helped improve the quality and presentation of the paper. We are grateful to Vithal Tilvi and James Rhoads for additional observations of SGAS1527 with NIRSPEC. We thank Wiphu Rujopakarn and George Rieke for sharing their new conversions between 24~$\mu$m emission and IR luminosity for high redshift galaxies in advance of publication. We thank Brian Siana for helpful discussions. EW acknowledges support from the Carnegie-Brinson Predoctoral Fellowship. Data presented in this paper were partly obtained at the W.M. Keck Observatory from telescope time allocated to the National Aeronautics and Space Administration through the scientific partnership with the California Institute of Technology and the University of California. The Observatory was made possible by the generous financial support of the W.M. Keck Foundation. We acknowledge the very significant cultural role and reverence that the summit of Mauna Kea has always had within the indigenous Hawaiian community. We are most fortunate to have the opportunity to conduct observations from this mountain.
\end{acknowledgments}


\begin{thebibliography}{}

\bibitem[Adelberger et al.(2004)]{Adelberger:04}
{Adelberger}, K.~L., {Steidel}, C.~C., {Shapley}, A.~E., {Hunt}, M.~P., {Erb}, D.~K., {Reddy}, N.~A.,\&  {Pettini}, M. 2004, \apj, 607, 226

\bibitem[Allam et al.(2007)]{Allam:07}
{Allam}, S.~S., {Tucker}, D.~L., {Lin}, H., {Diehl}, H.~T., {Annis}, J., {Buckley-Geer}, E.~J., \& {Frieman}, J.~A. 2007, \apjl, 662, L51

\bibitem[Anders \& Grevesse(1989)]{Anders:89}  
Anders, E., \& Grevesse, N. 1989, \gca, 53, 197 

\bibitem[Asplund et al.(2009)]{Asplund:09}
{Asplund}, M., {Grevesse}, N., {Sauval}, A. J., \& {Scott}, P. 2009, \araa, 47, 481

\bibitem[Baker et al.(2001)]{Baker:01}
Baker, A. J., Lutz, D., Genzel, R., Tacconi, L. J., \& Lehnert, M. D.  2001, \aap, 372, L37

\bibitem[Bayliss et al.(2011)]{Bayliss:11}
{Bayliss}, M.~B., {Hennawi}, J.~F., {Gladders}, M.~D., {Koester}, B.~P., {Sharon}, K., {Dahle}, H., \& {Oguri}, M. 2011, \apjs, 193, 8

\bibitem[Bell et al.(2005)]{Bell:05}
{Bell}, E.~F. et al. 2005, \apj, 625, 23

\bibitem[Belokurov et al.(2007)]{Belokurov:07}
{Belokurov}, V. et al. 2007, \apjl, 671, L9

\bibitem[Bertin \& Arnouts(1996)]{sex}
{Bertin}, E. \& {Arnouts}, S. 1996, \aaps, 117, 393

\bibitem[Bian et al.(2010)]{Bian:10}
{Bian}, F. et al. 2010, \apj, 725, 1877

\bibitem[Bolton et al.(2006)]{Bolton:06}
{Bolton}, A.~S., {Burles}, S., {Koopmans}, L.~V.~E., {Treu}, T., \& {Moustakas}, L.~A. 2006, \apj, 638, 703

\bibitem[Bolzonella et al.(2000)]{Bolzonella:00}
{Bolzonella}, M., {Miralles}, {J.-M.}, {Pell{\'o}}, R. 2000, \aap, 363, 476

\bibitem[Brinchmann et al.(2004)]{Brinchmann:04}
{Brinchmann}, J., {Charlot}, S., {White}, S.~D.~M., {Tremonti}, C., {Kauffmann}, G., {Heckman}, T.,\&  {Brinkmann}, J. 2004, \mnras, 351, 1151

\bibitem[Bruzual \& Charlot(2003)]{Bruzual:03}
{Bruzual}, G., {Charlot}, S. 2003, \mnras, 344, 1000

\bibitem[Buat et al.(2007)]{Buat:07}
{Buat}, V. 2007, \apjs, 173, 404

\bibitem[Cabanac et al.(2005)]{Cabanac:05}
{Cabanac}, R.~A., {Valls-Gabaud}, D., {Jaunsen}, A.~O., {Lidman}, C., \& Jerjen, H. 2005, \aap, 436, L21

\bibitem[Cabanac et al.(2007)]{Cabanac:07}
{Cabanac}, R.~A., et al. 2007, \aap, 461, 813

\bibitem[Calzetti et al.(1994)]{Calzetti:94}
{Calzetti}, D., {Kinney}, A.~L., \& {Storchi-Bergmann}, T. 1994, \apj, 429, 582

\bibitem[Calzetti et al.(2000)]{Calzetti:00}
{Calzetti}, D., {Armus}, L., {Bohlin}, R.~., {Kinney}, A.~L., {Koornneef}, J., \& {Storchi-Bergmann}, T. 2000, \apj, 533, 682

\bibitem[Calzetti et al.(2005)]{Calzetti:05}
{Calzetti}, D. et al. 2005, \apj, 633, 871

\bibitem[Chabrier(2003)]{Chabrier:03}
{Chabrier}, G. 2003, \pasp, 115, 763


\bibitem[Chary \& Elbaz(2001)]{Chary:01}
Chary, R., \& Elbaz, D. 2001, \apj, 556, 562

\bibitem[Daddi et al.(2004)]{Daddi:04}
{Daddi}, E. et al. 2004, \apj, 617, 746

\bibitem[Daddi et al.(2007)]{Daddi:07}
{Daddi}, E. 2007, \apj, 670, 156

\bibitem[Dekel et al.(2009)]{Dekel:09}
{Dekel}, A. 2009, \nat, 457, 451

\bibitem[Egami et al.(2010)]{Egami:10}
{Egami}, E. 2010, \aap, 518, L12

\bibitem[Elbaz et al.(2010)]{Elbaz:10}
{Elbaz}, D. 2001, \aap, 518, L29

\bibitem[Ellingson et al.(1996)]{Ellingson:96}
{Ellingson}, E., {Yee}, H.~K.~C., {Bechtold}, J., \& {Elston}, R. 1996, \apjl, 466, L71

\bibitem[Erb et al.(2006a)]{Erb:06a}
{Erb}, D.~K., {Shapley}, A.~E., {Pettini}, M., {Steidel}, C.~C., {Reddy}, N.~A., {Adelberger}, K.~L. 2006a, \apj, 644, 813

\bibitem[Erb et al.(2006b)]{Erb:06b}
{Erb}, D.~K., {Steidel}, C.~C., {Shapley}, A.~E., {Pettini}, M., {Reddy}, N.~A., {Adelberger}, K.~L. 2006b, \apj, 646, 107

\bibitem[Erb et al.(2006c)]{Erb:06c}
{Erb}, D.~K., {Steidel}, C.~C., {Shapley}, A.~E., {Pettini}, M., {Reddy}, N.~A., {Adelberger}, K.~L. 2006c, \apj, 647, 128

\bibitem[Fazio et al.(2004)]{Fazio:04}
{Fazio}, G.~G. et al. 2004, \apjs, 154, 10

\bibitem[Finkelstein et al.(2009)]{Finkelstein:09}
{Finkelstein}, S.~L. et al. 2009, \apj, 700, 376

\bibitem[F{\"o}rster Schreiber et al.(2009)]{Forsterschreiber:09}
{F{\"o}rster Schreiber}, N.~M. et al. 2009, \apj, 706, 1364

\bibitem[Gilbank et al.(2010)]{Gilbank:10}
{Gilbank}, D.~G., {Baldry}, I.~K., {Balogh}, M.~L., {Glazebrook}, K.,\&  {Bower}, R.~G. 2010, \mnras, 405, 2594

\bibitem[Gilbank et al.(2011)]{Gilbank:11}
{Gilbank}, D.~G., {Gladders}, M.~D., {Yee}, H.~K.~C.,\& {Hsieh}, B.~C. 2011, \aj, 141, 94

\bibitem[Gonzaga \& Biretta et al.(2010)]{wfpc2dhb}
{Gonzaga}, S., {Biretta}, J.  et al. 2010, HST WFPC2 Data Handbook, v. 5.0, ed., Baltimore, STScI

\bibitem[Hainline et al.(2009)]{Hainline:09}
{Hainline}, K.~N., {Shapley}, A.~E., {Kornei}, K.~A., {Pettini}, M., {Buckley-Geer}, E., {Allam}, S.~S., \& {Tucker}, D.~L. 2009, \apj, 701, 52

\bibitem[Hennawi et al.(2008)]{Hennawi:08}
{Hennawi}, J.~F. et al. 2008, \aj, 135, 664

\bibitem[Kennicutt(1998)]{Kennicutt:98}
{Kennicutt}, Jr., R.~C. 1998, \araa, 36, 189

\bibitem[Kewley \& Dopita(2002)]{Kewley:02} 
Kewley, L.~J., \& Dopita, M.~A. 2002, \apjs, 142, 35 

\bibitem[Kim \& Koo(2001)]{KimKoo:01} 
{Kim}, {K.-T.}, \& {Koo}, {B.-C.} 2001, \apj, 549, 979
 
\bibitem[Koester et al.(2010)]{Koester:10}
{Koester}, B.~P. et al. 2010, \apjl, 723, L73

\bibitem[Law et al.(2007)]{Law:07}
{Law}, D.~R. et al. 2007, \apj, 656, 1

\bibitem[Lilly et al.(2003)]{Lilly:03}
{Lilly}, S.~J., {Carollo}, C.~M., \& {Stockton}, A.~N. 2003, \apj, 597, 730
 
\bibitem[Lin et al.(2009)]{Lin:09}
{Lin}, H. et al. 2009, \apj, 699, 1242

\bibitem[Madau(1995)]{Madau:95}
{Madau}, P. 1995, \apj, 441, 18

\bibitem[Mannucci et al.(2009)]{Mannucci:09}
{Mannucci}, F. et al. 2009, \mnras, 398, 1915

\bibitem[Maraston et al.(2010)]{Maraston:10}
{Maraston}, C., {Pforr}, J., {Renzini}, A., {Daddi}, E., {Dickinson}, M., {Cimatti}, A., \& {Tonini}, C. 2010, \mnras, 407, 830

\bibitem[Marchesini et al.(2009)]{Marchesini:09}
{Marchesini}, D., {van Dokkum}, P.~G., {F{\"o}rster Schreiber}, N.~M., {Franx}, M., {Labb{\'e}}, I.,\&  {Wuyts}, S. 2009, \apj, 701, 1765

\bibitem[McLean et al.(1998)]{McLean:98}
{McLean}, I.~S. et al. 1998, Society of Photo-Optical Instrumentation Engineers (SPIE) Conference Series, 3354, 566
 
\bibitem[Meurer et al.(1999)]{Meurer:99}
{Meurer}, G.~R., {Heckman}, T.~M., \& {Calzetti}, D. 1999, \apj, 521, 64

\bibitem[Moustakas et al.(2006)]{Moustakas:06}
{Moustakas}, J., {Kennicutt}, Jr., R.~C., \& {Tremonti}, C.~A. 2006, \apj, 642, 775

\bibitem[Moustakas et al.(2010)]{Moustakas:10}
{Moustakas}, J., {Kennicutt}, Jr., R.~C., {Tremonti}, C.~A., {Dale}, D.~A., {Smith}, {J.-D.~T.}, \& {Calzetti}, D. 2010, \apjs, 190, 233

\bibitem[Noeske et al.(2007)]{Noeske:07}
{Noeske}, K.~G. et al. 2007, \apjl, 660, L43

\bibitem[Nordon et al.(2010)]{Nordon:10}
{Nordon}, R. 2001, \aap, 518, L24

\bibitem[Oesch et al.(2010)]{Oesch:10}
{Oesch}, P.~A. et al. 2010, \apjl, 725, L150

\bibitem[Osterbrock(1989)]{Osterbrock:89}
{Osterbrock}, D.~E. 1989, Astrophysics in Gaseous Nebulae and Active Galactic Nuclei

\bibitem[Panella et al.(2009)]{Panella:09}
Pannella, M. et al. 2009, \apj, 698, L116

\bibitem[Papovich et al.(2001)]{Papovich:01}
{Papovich}, C., {Dickinson}, M., {Ferguson}, H.~C. 2001, \apj, 559, 620

\bibitem[Papovich et al.(2007)]{Papovich:07}
{Papovich}, C. et al. 2007, \apj, 668, 45

\bibitem[Peng et al.(2010)]{Peng:10}
{Peng}, C.~Y., {Ho}, L.~C., {Impey}, C.~D., \& {Rix}, {H.-W.} 2010, \aj, 139, 2097

\bibitem[Prevot et al.(1984)]{Prevot:84}
{Prevot}, M.~L., {Lequeux}, J., {Prevot}, L., {Maurice}, E., \& {Rocca-Volmerange}, B. 1984, \aap, 132, 389

\bibitem[Reddy et al.(2006)]{Reddy:06}
{Reddy}, N.~A. et al. 2006, \apj, 644, 792

\bibitem[Reddy \& Steidel(2009)]{Reddy:09}
{Reddy}, N.~A., \&  {Steidel}, C.~C. 2009, \apj, 692, 778

\bibitem[Reddy et al.(2010)]{Reddy:10}
{Reddy}, N.~A., {Erb}, D.~K., {Pettini}, M., {Steidel}, C.~C., \& {Shapley}, A.~E. 2010, \apj, 712, 1070

\bibitem[Renzini \& Ciotti(1993)]{Renziniciotti:93}
{Renzini}, A., {Ciotti}, L. 1993, \apjl, 416, L49

\bibitem[Rex et al.(2010)]{Rex:10}
{Rex}, M. 2010, \aap, 518, L13

\bibitem[Rieke et al.(2004)]{Rieke:04}
{Rieke}, G.~H. et al. 2004, \apjs, 154, 25

\bibitem[Rieke et al.(2009)]{Rieke:09}
{Rieke}, G.~H., {Alonso-Herrero}, A., {Weiner}, B.~J., {P{\'e}rez-Gonz{\'a}lez}, P.~G., {Blaylock}, M., {Donley}, J.~L.,\&  {Marcillac}, D. 2009, \apj, 692, 556

\bibitem[Rigby et al.(2008)]{Rigby:08}
{Rigby}, J.~R. et al. 2008, \apj, 675, 262

\bibitem[Rigby et al.(2011)]{Rigby:11}
{Rigby}, J.~R., {Wuyts}, E., {Gladders}, M., {Sharon}, K.,\&  {Becker}, G.~D. 2011, \apj, 732, 59

\bibitem[Rujopakarn et al.(2011a)]{Rujopakarn:11a}
{Rujopakarn}, W., {Rieke}, G.~H., {Eisenstein}, D.~J., \& {Juneau}, S. 2011, \apj, 726, 93

\bibitem[Rujopakarn et al.(2011b)]{Rujopakarn:11b}
{Rujopakarn}, W., {Rieke}, G.~H., {Weiner}, B.~J., {Rex}, M., {Walth}, G.~L., \& {Kartaltepe}, J.~S. 2011, arXiv1107.2921

\bibitem[Salim et al.(2007)]{Salim:07}
{Salim}, S. et al. 2007, \apjs, 173, 267

\bibitem[Schlegel (1998)]{Schlegel:98}
{Schlegel}, D.~J., {Finkbeiner}, D.~P.,\&  {Davis}, M. 1998, \apj, 500, 525

\bibitem[Seaton(1979)]{Seaton:79}
{Seaton}, M.~J. 1979, \mnras, 187, 73P

\bibitem[Seitz et al.(1998)]{Seitz:98}
{Seitz}, S., {Saglia}, R.~P., {Bender}, R., {Hopp}, U., {Belloni}, P.,\&  {Ziegler}, B. 1998, \mnras, 298, 945

\bibitem[Shapley et al.(2001)]{Shapley:01}
{Shapley}, A.~E., {Steidel}, C.~C., {Adelberger}, K.~L., {Dickinson}, M., {Giavalisco}, M., \& {Pettini}, M. 2001, \apj, 562, 95

\bibitem[Shapley et al.(2005)]{Shapley:05}
{Shapley}, A.~E. et al. 2005, \apj, 626, 698

\bibitem[Sharon et al.(2011)]{Sharon:11}
{Sharon}, K., Gladders, M.~D., Rigby, J.~R., Wuyts, E., Koester, B.~P., Bayliss, M.~B., Barrientos, L.~F. 2011, in preparation

\bibitem[Shi et al.(2007)]{Shi:07}  
Shi, F., Zhao, G., \& Liang, Y.~C. 2007, \aap, 475, 409 

\bibitem[Siana et al.(2008)]{Siana:08}
{Siana}, B., {Teplitz}, H.~I., {Chary}, {R.-R.}, {Colbert}, J., \& {Frayer}, D.~T. 2008, \apj, 689, 59

\bibitem[Siana et al.(2009)]{Siana:09}
{Siana}, B. et al. 2009, \apj, 698, 1273

\bibitem[Smail et al.(2007)]{Smail:07}
{Smail}, I. et al. 2007, \apjl, 654, L33

\bibitem[Steidel et al.(2004)]{Steidel:04}
{Steidel}, C.~C., {Shapley}, A.~E., {Pettini}, M., {Adelberger}, K.~L., {Erb}, D.~K., {Reddy}, N.~A., \& {Hunt}, M.~P. 2004, \apj, 604, 534

\bibitem[Teplitz et al.(2000)]{Teplitz:00}
{Teplitz}, H.~I. et al. 2000, \apjl, 533, L65

\bibitem[Tremonti et al.(2004)]{Tremonti:04}
{Tremonti}, C.~A. et al. 2004, \apj, 613, 898 

\bibitem[van der Burg et al.(2010)]{vanderBurg:10}
{van der Burg}, R.~F.~J., {Hildebrandt}, H., \& {Erben}, T. 2010, \aap, 523, A74

\bibitem[van der Werf et al.(2001)]{vanderwerf:01}
{van der Werf}, P.~P., {Knudsen}, K.~K., {Labb{\'e}}, I., \& {Franx}, M. 2001, Deep Millimeter Surveys: Implications for Galaxy Formation and Evolution, ed. J. D. Lowenthal \& D. H. Hughes, 103–+

\bibitem[Wuyts et al.(2010)]{Wuyts:10}
{Wuyts}, E. et al. 2010, \apj, 724, 1182

\bibitem[Wuyts et al.(2007)]{Wuyts:07}
{Wuyts}, S. et al. 2007, \apj, 655, 51

\bibitem[Wuyts et al.(2011)]{Wuyts:11}
{Wuyts}, S. et al. 2011, arXiv1106.5502

\bibitem[Yee et al.(1996)]{Yee:96}
{Yee}, H.~K.~C., {Ellingson}, E., {Bechtold}, J., {Carlberg}, R.~G., {Cuillandre}, {J.-C.} 1996, \aj, 111, 1783

\bibitem[Zaritsky et al.(1994)]{Zaritsky:94}
Zaritsky, D., Kennicutt, R.~C., \& Huchra, J.~P. 1994, \apj, 420, 87

\end{thebibliography}
\end{document}